\documentclass[11pt]{article}
\pdfoutput=1
\usepackage{jcappub}

\usepackage[T1]{fontenc} 
\usepackage{verbatim}
\usepackage[utf8]{inputenc}
\usepackage[american]{babel}
\usepackage{graphicx}
\usepackage{epsfig}
\usepackage{adjustbox}
\usepackage{booktabs}
\usepackage{multirow}
\usepackage{dcolumn}
\usepackage{amsmath}
\usepackage{amsfonts}
\usepackage{mathtools}
\usepackage{amssymb}
\usepackage{subcaption} 
\usepackage{caption}
\usepackage{euscript,bbm,amsthm}
\usepackage{ifthen}
\usepackage{epstopdf}
\usepackage{bm}
\usepackage{mathrsfs}
\usepackage{version}
\usepackage{enumerate}
\usepackage{braket}
\usepackage{enumitem}
\usepackage{transparent}
\usepackage{pifont}
\usepackage{colortbl}
\usepackage{rotating}
\usepackage{adjustbox}
\usepackage{cancel}
\usepackage{tabularx}
\usepackage{bbold}
\usepackage{soul}
\usepackage[table]{xcolor}
\usepackage{siunitx}
\usepackage{float}
\usepackage{cite}

\usepackage{hyperref}
\hypersetup{colorlinks=true,linkcolor=royalblue,citecolor=magenta,
linkcolor=navyblue,citecolor=orange(colorwheel),urlcolor=magenta(process),pdfencoding=auto}

\definecolor{navyblue}{rgb}{0.0, 0.0, 0.5}
\definecolor{royalblue}{rgb}{0.25, 0.41, 0.88}
\definecolor{cadmiumgreen}{rgb}{0.0, 0.42, 0.24}
\definecolor{blue-violet}{rgb}{0.54, 0.17, 0.89}
\definecolor{darkviolet}{rgb}{0.58, 0.0, 0.83}
\definecolor{orange(colorwheel)}{rgb}{1.0, 0.5, 0.0}

\definecolor{magenta(process)}{rgb}{1.0, 0.0, 0.56}

\definecolor{darkspringgreen}{rgb}{0.09, 0.45, 0.27}

\definecolor{royalblue(web)}{rgb}{0.25, 0.41, 0.88}

\definecolor{cadmiumorange}{rgb}{0.93, 0.53, 0.18}

\definecolor{heliotrope}{rgb}{0.87, 0.45, 1.0}

\makeatletter
\renewcommand*{\@textcolor}[3]{%
\protect\leavevmode
\begingroup
\color#1{#2}#3%
\endgroup
}
\makeatother

\renewcommand\[{\left[}

\DeclarePairedDelimiter{\abs}{\lvert}{\rvert}

\makeatletter
\let\save@mathaccent\mathaccent
\newcommand*\if@single[3]{%
\setbox0\hbox{${\mathaccent"0362{#1}}^H$}%
\setbox2\hbox{${\mathaccent"0362{\kern0pt#1}}^H$}%
\ifdim\ht0=\ht2 #3\else #2\fi
}
\newcommand*\rel@kern[1]{\kern#1\dimexpr\macc@kerna}
\newcommand*\widebar[1]{\@ifnextchar^{{\wide@bar{#1}{0}}}{\wide@bar{#1}{1}}}
\newcommand*\wide@bar[2]{\if@single{#1}{\wide@bar@{#1}{#2}{1}}{\wide@bar@{#1}{#2}{2}}}
\newcommand*\wide@bar@[3]{%
\begingroup
\def\mathaccent##1##2{%
\let\mathaccent\save@mathaccent
\if#32 \let\macc@nucleus\first@char \fi
\setbox\z@\hbox{$\macc@style{\macc@nucleus}_{}$}%
\setbox\tw@\hbox{$\macc@style{\macc@nucleus}{}_{}$}%
\dimen@\wd\tw@
\advance\dimen@-\wd\z@
\divide\dimen@ 3
\@tempdima\wd\tw@
\advance\@tempdima-\scriptspace
\divide\@tempdima 10
\advance\dimen@-\@tempdima
\ifdim\dimen@>\z@ \dimen@0pt\fi
\rel@kern{0.6}\kern-\dimen@
\if#31
\overline{\rel@kern{-0.6}\kern\dimen@\macc@nucleus\rel@kern{0.4}\kern\dimen@}%
\advance\dimen@0.4\dimexpr\macc@kerna
\let\final@kern#2%
\ifdim\dimen@<\z@ \let\final@kern1\fi
\if\final@kern1 \kern-\dimen@\fi
\else
\overline{\rel@kern{-0.6}\kern\dimen@#1}%
\fi
}%
\macc@depth\@ne
\let\math@bgroup\@empty \let\math@egroup\macc@set@skewchar
\mathsurround\z@ \frozen@everymath{\mathgroup\macc@group\relax}%
\macc@set@skewchar\relax
\let\mathaccentV\macc@nested@a
\if#31
\macc@nested@a\relax111{#1}%
\else
\def\gobble@till@marker##1\endmarker{}%
\futurelet\first@char\gobble@till@marker#1\endmarker
\ifcat\noexpand\first@char A\else
\def\first@char{}%
\fi
\macc@nested@a\relax111{\first@char}%
\fi
\endgroup
}
\makeatother

\renewcommand{\bar}{\widebar}

\newcommand{\ex}[1]{\left\langle #1 \right\rangle}

\newcommand\ee{\end{eqnarray}}
\newcommand\be{\begin{eqnarray}}
\newcommand{\bsp}{\begin{split}}
\newcommand{\esp}{\end{split}}
\newcommand{\bit}{\begin{itemize}[leftmargin=*]}
\newcommand{\eit}{\end{itemize}}
\newcommand{\ben}{\begin{enumerate}[leftmargin=*]}
\newcommand{\een}{\end{enumerate}}

\newcommand{\ie}{\textit{i.e.}~}
\newcommand{\eg}{\textit{e.g.}~}

\newcommand{\mpc}[1]{\SI{#1}{\mathrm{Mpc}^{-1}}}

\newcommand\eq[1]{Eq.~\eqref{eq:#1}}
\newcommand\eqsI[1]{Eqs.~\eqref{eq:#1}}
\newcommand{\eqsII}[2]{Eqs.~\eqref{eq:#1}, \eqref{eq:#2}}
\newcommand{\eqsIII}[3]{Eqs.~\eqref{eq:#1}, \eqref{eq:#2}, \eqref{eq:#3}}

\newcommand{\dif}{\mathrm{d}}

\renewcommand{\vec}{\bm} 
\newcommand\vers[1]{\hat{\vec{#1}}}

\newcommand{\cH}{\mathcal{H}}

\def\mpl{M_{\rm Pl}}

\def\qD{q_{\mu,\mathrm{D}}}
\def\fnl{f_{\rm NL}}

\begin{document}

\begin{titlepage}

\setcounter{page}{1} \baselineskip=15.5pt \thispagestyle{empty}

\bigskip\

\vspace{1cm}
\begin{center}


{
\huge{\textbf{\textsf{Spectral distortion anisotropies\\[10pt] from single-field inflation}}}}

\end{center}

\vspace{0.2cm}

\begin{center}
{
\Large{Giovanni Cabass,$^{\rm a}$ Enrico Pajer,$^{\rm b}$ and Drian van der Woude$^{\rm b}$}}
\end{center}

\begin{center}

\textsl{\small{
$^{\rm a}$ Max-Planck-Institut f\"{u}r Astrophysik, Karl-Schwarzschild-Str. 1, 85741 Garching, Germany
\\[0.25cm]
$^{\rm b}$ Institute for Theoretical Physics and Center for Extreme Matter and Emergent Phenomena, \\
Utrecht University, 
Princetonplein 5, 3584 CC Utrecht, The Netherlands
}}
\vskip 7pt

\end{center}

\vspace{1.2cm}
\hrule \vspace{0.3cm}
\noindent {
\textbf{\textsf{Abstract}}} \\[0.1cm]
Distortions of the Cosmic Microwave Background energy spectrum of the $\mu$ type are sensitive to the primordial power spectrum through 
the dissipation of curvature perturbations on scales $k\simeq 50$ - $10^4\,\mathrm{Mpc}^{-1}$. 
Their angular correlation with large-scale temperature anisotropies is then sensitive to the squeezed limit of the primordial bispectrum. 
For inflationary models obeying the single-field consistency relation, 
we show that the observed $\mu T$ angular correlation that would correspond to the local shape vanishes exactly. 
All leading non-primordial contributions, including all non-linear production and projection effects, are of the ``equilateral shape'', 
namely suppressed by $k^2/\cH_f^2$, where $\cH_f\simeq\mpc{e-1}$ is the Hubble radius at the end of the $\mu$-era. 
Therefore, these non-primordial contributions are orthogonal to a potential local primordial signal (\textit{e.g.} from multi-field inflation). 
Moreover, they are very small in amplitude. 
Our results strengthen the position of $\mu$ distortions as the ultimate probe of local primordial non-Gaussianity. 

\vskip 10pt
\hrule

\vspace{0.6cm}
\end{titlepage}

\tableofcontents

\clearpage 
\flushbottom

\section{Introduction and main results}
\label{sec:introduction}

\noindent As it is well-known \cite{Hu:1994bz}, $\mu$-type distortions of the Cosmic Microwave Background (CMB) energy spectrum probe primordial perturbations 
on scales $k\simeq 50$ - $10^4\,\mathrm{Mpc}^{-1}$ through the dissipation of acoustic waves in the photon-electron-baryon fluid 
(for a recent review see \cite{Chluba:2015bqa} and references therein). 
In presence of primordial non-Gaussianity, 
the amplitude of the dissipation becomes spatially dependent on large scales and it gives rise to an angular correlation $C^{\mu T}_\ell$ between 
$\mu(\hat{\vec{n}})$ and large-scale $T(\hat{\vec{n}})$ anisotropies, 
which are sourced by the same large-scale modes that modulate the dissipation \cite{Pajer:2012vz}. 
Indeed, with local non-Gaussianity we expect an angular cross-correlation 
between temperature anisotropies and fractional $\mu$ anisotropies given by 
${-12}f_\mathrm{NL} C^{TT}_\ell$ \cite{Pajer:2012vz,Emami:2015xqa}.\footnote{We will consider only local $f_\text{NL}$ in this work. 
Conventionally, it is defined in terms of the Newtonian potential during matter domination, 
\ie $B_\Phi(k_1,k_2,k_3)={-2\fnl} P_\Phi(k_1)P_\Phi(k_2)+\text{$2$ perms.}$, where $\zeta = {-\frac{5\Phi}{3}}$ \cite{Komatsu:2001rj,Bartolo:2004if}. 
Notice also that we use the notation of \cite{Maldacena:2002vr} for the comoving curvature perturbation. } 
Much recent work has been devoted to better understand and model this mechanism as well as forecasting 
and measuring the constraining power of $C^{\mu T}_\ell$ for primordial non-Gaussianity 
\cite{Ganc:2012ae,Pajer:2013oca,Ota:2014iva,Naruko:2015pva,
Khatri:2015tla,Emami:2015xqa,Bartolo:2015fqz,Shiraishi:2016hjd,
Dimastrogiovanni:2016aul,Chluba:2016aln,Ota:2016mqd,Ravenni:2017lgw,Remazeilles:2018kqd}. 

As for CMB temperature anisotropies, 
one expects late-time evolution to ``contaminate'' any contribution from primordial non-Gaussianity. 
For example, even for single-field inflation satisfying the Maldacena's consistency relation
there is a non-primordial contribution to the squeezed CMB bispectrum of the local shape, $B^{TTT}_{\ell_L\ell_S\ell_S}\sim C^{TT}_{\ell_L}C^{TT}_{\ell_S}$
\cite{Bartolo:2003gh,Bartolo:2003bz,Creminelli:2004pv,Creminelli:2011sq,Bartolo:2011wb,Lewis:2012tc,Jeong:2013psa,Pajer:2013ana}. 
For $\ell_L\lesssim 100$, the long mode is outside the Hubble radius at recombination and therefore it cannot change the local physics to 
leading and subleading order in derivatives. 
Instead, the long mode affects photons as they travel from the last-scattering surface to the observer. 
For single-field inflation models that satisfy the consistency relation, this effect is all the result \cite{Creminelli:2011sq,Pajer:2013ana,Mirbabayi:2014hda}. 

In this paper, we show that an analogous \textit{non-primordial contamination is instead absent for the $C^{\mu T}_\ell$ angular spectrum:
in single-field attractor inflation $C^{\mu T}_\ell$ vanishes up to corrections of order $k^2/\mathcal{H}_f^2$}, 
where $k\sim \ell/\eta_0$ is the long-wavelength temperature mode and 
$\mathcal{H}_f\simeq\mpc{e-1}$ is the Hubble radius at the end of the so-called $\mu$-era, $z\simeq5\times10^4$, 
when $\mu$ distortions stop being generated. 
Non-primordial contributions to $ \mu T $ arise from non-linear ``production'' effects, \textit{i.e.} non-linearities during the $ \mu $-era, 
and from non-linear ``propagation'' effects, \textit{i.e.} non-linearities in the evolution from the $ \mu $-era to observation. 
We show that the leading non-primordial contributions lead to a $C^{\mu T}_\ell$ that has the same $ \ell $-dependence as equilateral non-Gaussianity, 
as opposed to local non-Gaussianity. In addition, the amplitude of these contributions is very small 
and would be detectable only by a very futuristic almost cosmic variance-limited experiment 
(see Section \ref{sec:forecast} for details). 

Our results can be intuitively understood as follows. To compute $C^{\mu T}_\ell$, we can divide the sky in patches,
measure the average chemical potential $ \mu(\hat{\vec{n}}) $ in each patch, and then see if it correlates with $ T(\hat{\vec{n}}) $. 
All modes that are observationally relevant were super-Hubble during the $ \mu $-era, 
and therefore their effect on the production of $ \mu $ (which is a local observable) were suppressed by at least two derivatives over the Hubble radius. 
Therefore, these non-primordial production effects lead to an effective equilateral shape 
of $C^{\mu T}_\ell$, as opposed to local.\footnote{We use the loose language of ``equilateral shape'' 
and ``local shape'' to refer to the $ \ell $ dependence of $C^{\mu T}_\ell$ that would be generated by equilateral or local primordial non-Gaussianity.} 
No spatial variation of $ \mu $ can come from the initial conditions either, 
if the consistency relation of single-clock inflation is satisfied \cite{Pajer:2013ana}. 
We see, then, that there are only two effects that could contribute to $ C^{\mu T}_{\ell} $, 
both coming from so-called ``projection effects'' as the photons travel to us:
\begin{itemize}
\item The same physical length scale appears at different angular sizes to the observer, 
due to the expansion and distortion of the photon geodesics caused by a long mode.
\item Photons experience a different redshift in different directions due to the presence of a long mode.
\end{itemize}
The modulation of physical scales does not lead to any effect when we compute $C^{\mu T}_\ell$ 
because the average $ \mu $ in a patch of the sky does not posses any intrinsic length scale. 
Let us contrast this with the temperature bispectrum. There, what we are doing is measure the $TT$ angular power spectrum in 
each patch and check if it correlates with a long temperature mode. In this case, we do have a physical length scale that can be distorted by the long mode, 
namely to the distance between the peaks of the short-scale temperature power spectrum. 
When we consider the $\mu T$ power spectrum, instead, we are just looking at the average $\mu$ in a patch, 
so there is no ``ruler'' whose length the long mode can perturb. 
In other words, a homogeneous $ \mu(\vec{x})=\mu $ remains such under evolution after the $ \mu $-era 
because only inhomogeneities in $ \mu(\vec{x}) $ can be lensed or deformed. 
The second effect also vanishes for $ C^{\mu T}_{\ell} $. To see this, let us contrast it again with the temperature bispectrum. 
In that case, a long mode modifies the Sachs-Wolfe + Doppler + Integrated Sachs-Wolfe formula relating 
inhomogeneities in the photon distribution at recombination to temperature anisotropies at the observer's point, 
and changes the average temperature in each patch (temperature with respect to which the anisotropies are defined). 
Instead, since $\mu$ does not depend on the photon energy, 
we can and do always define it to be the dimensionless quantity $\mu = -\mu_\text{th}/T$, 
where $\mu_\text{th}$ has its usual definition from thermodynamics. 
This variable does not redshift: the evolution after the $ \mu $-era leaves it unchanged. 

Finally, in addition to showing that the non-primordial contributions to $\mu T$ cross-correlations are of all of the equilateral shape and small in amplitude, 
we also derive two new, albeit more technical, results:
\begin{itemize}
\item We rewrite the hydrodynamic description of \cite{Pajer:2013oca} in a manifestly covariant formalism. 
This allows us to write simple non-perturbative expressions for the generation of $\mu$ distortions, such as Eq.~\eqref{eq:mu_generation-B}. 
\item We derive for the first time a Fourier window function for the generation of $\mu$ distortions at second order in perturbations 
that is valid for modes of all wavelengths. This is to be contrasted with the expressions used in the literature that are valid only in the sub-Hubble regime. 
Although the dissipation does indeed mostly come from sub-Hubble modes, 
our general-relativistic formula is nevertheless essential to correctly compute 
one of the contributions to the $\braket{\mu T}$ correlator beyond the result discussed above, 
and prove that it is subleading with respect to it. 
\end{itemize}

\paragraph{Outline} The outline of the paper is as follows. 
In Section \ref{sec:general} we describe the general strategy of the computation and review our assumptions. 
In Section \ref{sec:mu_era} we review the creation of $\mu$ distortions from Silk damping, \textit{i.e.} by viscosity, and 
estimate the effect of the long-wavelength temperature mode on $\mu$ production. 
In Section \ref{sec:after_mu_era} we discuss the evolution of $\mu$ from the end of the $\mu$-era to the observer following the approach of \cite{Creminelli:2011sq}, 
and show that no correlation with the long mode is generated during this time. 
In Section \ref{sec:leading_effect}, then, we compute the 
leading non-primordial effect on the observed $\mu T$ correlator, which comes from the modulation of the $\mu$ production from Silk damping 
by the long-wavelength temperature mode. In this Section we also discuss other subleading contributions, \ie 
the effect of temperature non-linearities and of heat conduction. 
Finally we show that these non-primordial contributions are orthogonal to those of a local $\fnl$. 
Conclusions are drawn in Section \ref{sec:conclusions}. 
The Appendices from \ref{app:appendix-1-A} to \ref{app:forecast} contain some technical details on the computations of 
Sections \ref{sec:mu_era}, \ref{sec:after_mu_era}, \mbox{and \ref{sec:leading_effect}.} 
The Appendix \ref{app:PIXIE} confirms that also for a PIXIE-like experiment the impact of the non-primordial contributions on 
$\sigma(\fnl)$ will be negligible.


\section{General strategy and assumptions}
\label{sec:general}

\noindent In this section, we outline our general strategy for the calculation of the $\mu T$ cross-correlation in single-field inflation. 
Then, for the convenience of the reader, we summarize and discuss the main assumptions and approximations we will employ.


\subsection{Preliminaries}
\label{ssec:preliminaries}

\noindent Our goal is to compute the largest contribution to the observable $\mu T$ cross-correlation at late times. 
For multi-field models with sizable \textit{local} primordial non-Gaussianity, $ \fnl\gg \mathcal{O}(n_{\rm s}-1) $, 
the largest contribution is proportional to $ \fnl $ and was first computed in \cite{Pajer:2012vz}. 
Instead, as anticipated in \cite{Pajer:2012vz}, for single-field inflation the well-known contribution $ \fnl=(1-n_{\rm s}) $ is a gauge artifact 
and should cancel exactly in the final observable result. 
Here, besides showing this more explicitly, 
we establish that \textit{the next largest surviving contribution to $ \braket{\mu T} $ is the non-linear evolution of short modes during the $\mu$-era, 
which feel a long mode as a local spatial curvature}. 
The double derivative suppression of this non-linear effect leads to a final contribution to $ \mu T $ that is of the equilateral shape and small in amplitude.

Let us introduce and define some quantities of interest. We expand the perturbations in $\mu$ and $T$ as 
\begin{subequations}
\begin{align}
&\mu= \mu^{(1)}+\mu^{(2)}+\mu^{(3)}+\dots\,\,, \\
&T=T^{(1)}+T^{(2)}+\dots\,\,.
\end{align}
\end{subequations}
Note that $\mu$ contains a linear contribution $ \mu^{(1)}\sim\mathcal{O}(\zeta) $ due to heat conduction \cite{Pajer:2013oca}. 
An additional linear contribution comes from perturbations to adiabatic cooling. 
This contribution is suppressed by the baryon-to-photon number ratio and turns out to be negligible. 
This is further discussed in Section \ref{ssec:assumptions}, around Eq.~\eqref{muAC}. 
Besides, as argued in \cite{Pajer:2013oca}, bulk viscosity effects are suppressed by the photon-to-baryon ratio squared and can be neglected. 
Then, the standard and largest contribution to $ \braket{\mu} $ starts at quadratic order, 
$\mu\sim\mathcal{O}(\zeta^{2})$. 
Also, we used the fact that we want to compute $\braket{\mu T} $ up to $\mathcal{O}(\zeta^{4})\sim\mathcal{O}(\Delta_\zeta^{4})$,\footnote{The term 
$ \braket{\mu^{(1)}T^{(3)}} $ is suppressed by $ \Delta^2_\zeta $ with respect to $ \braket{\mu^{(1)} T^{(1)}} $ and is therefore negligible.} 
with $\Delta^2_\zeta\simeq\num{2d-9}$ being the amplitude of primordial perturbations (on CMB scales). 
Expanding in perturbations one finds
\begin{equation}
\label{eq:G_NG}
\braket{\mu T}=\braket{\mu^{(2)}T^{(1)}}_{\text{NG}}+\braket{\mu^{(1)}T^{(1)}}_{\text{G}}+\braket{\mu^{(3)}T^{(1)}}_{\text{G}}+\braket{\mu^{(2)}T^{(2)}}_{\text{G}}+\dots\,\,,
\end{equation}
where the label ``$\mathrm{NG}$'' reminds us that $\braket{ \mu^{(2)} T^{(1)}}$ is proportional to the primordial bispectrum, while the other terms are not. 
Expressions for $T^{(1)}$, $T^{(2)}$ are known in the literature while $\mu^{(2)}$, $\mu^{(3)}$ are not fully known and so we need to compute them here:
\begin{itemize}
\item A sub-Hubble approximation for $\mu^{(2)} $ has been derived and used many times in the literature 
(most recently for example in \cite{Chluba:2016aln}). 
However, this sub-Hubble approximation is not sufficient for calculating the $\braket{\mu^{(2)}T^{(2)}}$ 
correlator on observationally relevant scales,\footnote{The sub-Hubble approximation is instead sufficient to compute $ \braket{\mu^{(2)}\mu^{(2)} } $ 
because the incorrect $ k $ scaling, when squared, 
makes the integral still correctly peak on sub-Hubble scales, where the window function is a good approximation of the exact result.} 
so in Section \ref{sec:mu_era} we derive a fully general-relativistic expression for $\mu^{(2)}$ that is valid at any scale, sub- and super-Hubble. 
\item The next-to-leading order contribution $\mu^{(3)}$ is not yet known. Here we estimate its leading term, 
for the purpose of computing $\braket{\mu^{(3)}T^{(1)}}$ on observationally relevant scales. 
\end{itemize} 

The calculation of $ \mu^{(3)} $ is the most technically challenging part of our work and will be 
presented in Sections \ref{sec:mu_era} and \ref{sec:after_mu_era}. 
To guide the reader, let us outline our general strategy.


\subsection{General strategy}
\label{ss:generalstrategy}

\noindent First of all, let us separate the evolution \textit{during} the $\mu$-era, \ie the period of time when $\mu$ can be created, 
from the evolution \textit{after} the $\mu$-era until now, when only an existing $\mu$ can be lensed or dissipated. 
These are discussed in Section \ref{sec:mu_era} and Section \ref{sec:after_mu_era}, respectively. 
We will prove that the only relevant non-linearities arise during the $\mu$-era. 

Since $\mu$ starts at $ \mathcal{O}(\zeta^{2}) $, computing $ \mu^{(3)} $ requires the knowledge of $ \zeta $ at second order, 
and in particular of the long-short mode coupling $ \zeta^{(2)}\sim \zeta^{(1)}_S\zeta^{(1)}_L $. 
This mode coupling is generated only when the short modes are inside the Hubble radius,\footnote{When the short modes are super-Hubble, 
they freeze out and no sizeable non-linearity is generated.} 
either during inflation or after Hubble re-entry during the $\mu$-era. 
The non-linearities during the inflationary and $\mu$-era can each be computed either in global coordinates or in local physical coordinates, 
a.k.a. Conformal Fermi Coordinates (CFC) \cite{Dai:2015rda,Dai:2015jaa}. 
The CFC calculation at zeroth and first order in $k_L$ gives a vanishing long-short mode coupling both during inflation and the $\mu$-era by construction. 
At $\mathcal{O}(k_{L}^{2})$ the inflationary period leads to slow-roll suppressed terms \cite{Cabass:2016cgp}
\begin{equation}
\zeta^{(2)}\supset 0.1\times (n_{\rm s}-1)\,\left( \frac{k_{L}}{k_{S}} \right)^{2}\,\zeta_{L}^{(1)}\zeta_{S}^{(1)} \qquad \text{(inflation, CFC)}\,\,.
\end{equation}
At the same order $ \mathcal{O}(k_{L}^{2}) $, the $\mu$-era induces non-linearities of order
\begin{equation}
\zeta^{(2)}\supset \left( \frac{k_{L}}{\cH} \right)^{2}\,\zeta_{L}^{(1)}\zeta_{S}^{(1)} \qquad \text{($\mu$-era, CFC)}\,\,.
\end{equation}
These can be understood as resulting from the evolution of short modes in the background of the long mode, 
which mimics a spatial curvature in the isotropic case. 
In Section \ref{mu3}, we argue that $ \cH $ should be the Hubble scale at the end of the $\mu$-era. 
Since it is mostly the dissipation of sub-Hubble modes that sources $\mu$, 
in the above formulae $ k_{S}\gg \cH $ and the non-linearities in the $\mu$-era are much larger than those during inflation, 
which can safely be neglected. 

Up to this point, we have been able to neglect the constant and gradient part of the long $\zeta$ mode by construction. 
This works as long as the long mode is super-Hubble. 
However, we wish to compute the evolution of the $\braket{\mu T}$ all the way up to today, 
including the effect of modes that were super-Hubble at the time of $\mu$ production, but are sub-Hubble now. 
The inclusion of these ``projection effects'' proceeds in two steps. 

First we have to include the constant and gradient long modes in our computation. 
As long as they are super-Hubble, this can be done at any preferred moment by Weinberg's adiabatic mode construction 
\cite{Weinberg:2003sw,Creminelli:2004pv,Creminelli:2011sq,Mirbabayi:2014hda}: 
the effect of the long modes on CFC computations is equivalent to a change of coordinates. 
For us, the most convenient time to include these modes is at the end of the $\mu$-era. 
In fact, as we show in Section \ref{sec:after_mu_era}, since the expectation value of $\mu$ at this time is both independent of space and time, 
the effect of the coordinate change is zero!

Subsequently, we need to evolve the distribution of photons from the end of the $\mu$-era to today. 
Once again, we prove that the expectation value of $\mu$ is untouched during the evolution. 
The intuition is that any physical effect, such as lensing or dissipation, does not affect a homogeneous $\mu$. 
Hence, no spatial variation of $ \mu $ can be generated if there is none to begin with. 
In conclusion, no contribution to the $ \mu T $ cross-correlation arises from the non-linear evolution after the $ \mu $-era.

Note that Maldacena's $(n_{\rm s}-1)$ from the derivative of the short scale power spectrum \cite{Maldacena:2002vr} appears nowhere in this computation. 
This is a consequence of our convenient choice of the time at which we change from local (CFC) to global coordinates. 
If one were to insist on using global coordinates at some earlier stage, $ (n_{\rm s}-1) $ would appear. 
In that case, however, one would have to compute $\mu$ production in global coordinates and for consistency 
this should to be done at cubic order $ \mu^{(3)} = \mathcal{O}(\zeta^{3}) $. 
This is further discussed in Appendix \ref{app:nsmin1}. 
Since the final prediction for $\braket{\mu T}$ is independent of the method we use to calculate it, 
we stick to the most straightforward strategy.


\subsection{Assumptions}
\label{ssec:assumptions}

\noindent Let us summarize and discuss the main approximations and assumptions in our analysis.

\paragraph{Hydrodynamic approximation} The full calculation of the CMB spectrum anisotropies at second order in perturbations 
would require solving the inhomogeneous collisional Boltzmann equation, which is undoubtedly a daunting task. 
On the other hand, the system is perturbative and well amenable to the use of effective theory techniques, 
such as the hydrodynamical approach developed in \cite{Pajer:2013oca}. 
Here we follow this analytical approach, which makes the physics transparent at the cost of a tiny error coming from the approximation of thermodynamical equilibrium. 
The only shortcoming of this approach is that the boundaries of the $\mu$ and $ y $ eras must be given as an added input, 
derived from the homogeneous solution of the Boltzmann equation. 
Using for example some simple analytic fits to detailed numerical simulations (see \textit{e.g.} \cite{Chluba:2013vsa}) 
one expects this to lead to a mistake at the percent level.

\paragraph{Single fluid approximation} In principle, one should keep track of five fluids: photons, Dark Matter, neutrinos, baryons and electrons. 
Neutrinos are still relativistic at the time of interest and free streaming. 
Therefore neutrinos inhomogeneities quickly decay on sub-Hubble scales. 
We approximate the neutrinos as a homogeneous fluid, which contributes only to the background evolution. 
Dark Matter on sub-Hubble scales has inhomogeneities that grow very little (logarithmically) 
during radiation domination and linearly in $ a(t) $ during matter domination. 
During the $\mu$-era, the energy density of Dark Matter is much smaller than that of the photons, 
$ \rho_{\rm DM}/\rho_{\gamma}< 0.1 $. Since the interaction with any other component is gravitational, 
Dark Matter inhomogeneities can also be neglected during the $\mu$-era. 
Instead, we keep all Dark Matter inhomogeneities at later times, after the $ \mu $-era, 
since they could be important for example for the lensing calculation. 
Baryons, electrons and photons are tightly coupled before recombination and so the respective fluid velocities are the same to a good approximation. 
We can therefore treat them as a single fluid with some common velocity $ U^{\mu} $ and with 
\begin{subequations}
\begin{align}
&\rho = \rho_{\gamma}+\rho_{\rm b}\simeq \rho_{\gamma}+m_{\rm p} n_{\rm b}\,\,, \\
&p = p_{\gamma}+p_{\rm b}\simeq p_{\gamma}\,\,,
\end{align}
\end{subequations}
where the labels $ \gamma $ and $ \rm b $ refer to photons and baryons respectively and $ m_{\rm p} $ is the proton mass. 
Here we neglected the small contribution due to electrons, 
$ m_{\rm e}/m_{\rm p}\simeq 5 \times 10^{-4} $, and neglected the effect of baryon temperature, 
which are suppressed by the baryon-to-photon number ratio $ r=n_{\rm b}/n_{\gamma}\simeq 5 \times 10^{-10}$. 

\paragraph{Adiabatic cooling} Even for an ideal fluid, there is some homogeneous $\mu$ production during the $\mu$-era due to adiabatic cooling, 
which is proportional to the photon-to-baryon number ratio $r$. 
This is estimated to be $\bar{\mu}_\text{AC}\simeq -2.7 \times 10^{-9}$ for the \textit{Planck} 2015 best-fit parameters 
(see \eg \cite{Cabass:2016giw,Chluba:2016bvg} and references therein) and therefore slightly smaller than the homogeneous contribution from Silk damping, 
$\langle \mu_{S} \rangle \simeq \num{2d-8}$.

One might worry that long modes also modulate $\bar{\mu}_\text{AC}$, which then contributes to $\braket{\mu T}$. 
Fortunately, the treatment of the effect of long modes on $\bar{\mu}_\text{AC}$ is analogous to the treatment of $\mu$ from Silk damping. 
During the $\mu$-era, the non-linearities are computed in CFC, meaning corrections are at most of the form 
\begin{equation}
\label{muAC}
\mu_\text{AC}^{(1)}\supset \left(\frac{k_{L}}{\mathcal{H}}\right)^{2}\bar{\mu}_\text{AC}\,\zeta_{L}^{(1)} \qquad \text{($\mu$-era, CFC)}\,\,.
\end{equation}
After the $\mu$-era, we compute the change to global coordinates and the subsequent evolution in exactly the same way, 
and since $\bar{\mu}_\text{AC}$ is also homogeneous and frozen, it is untouched throughout its evolution and no further correlation with the long modes is induced. 
If indeed $\bar{\mu}_\text{AC}\ll \braket{\mu_{S}}$ as suggested by current data, 
we expect Eq.~\eqref{muAC} to lead to an effect that is subdominant with respect to the term $\braket{\mu^{(3)}T^{(1)}}_{\text{G}}$ in \eq{G_NG} 
(which is discussed in Section \ref{ssec:mu_cubic} below).


\section{Generation of \texorpdfstring{$\mu$}{\textbackslash mu}-type spectral distortion}
\label{sec:mu_era}

\noindent In this section, we review the generation of $\mu$-type spectral distortion to leading (second) order in perturbations, 
following \cite{Pajer:2013oca}. 
The formulae in the literature (see \textit{e.g.} \cite{Hu:1994bz,Pajer:2012vz,Chluba:2016aln,Ravenni:2017lgw}) 
for $\mu$ from the dissipation of acoustic modes are valid only in the sub-Hubble regime, 
which is indeed where most of the effect comes from. 
Here, we derive a more general formula for $\mu$ that is instead valid for arbitrary long wavelength perturbations. 
Our main result, Eqs.~\eqref{eq:window_function-A} and \eqref{eq:window_function-B}, 
can be thought of as the general-relativistic extension of the Fourier space window function used in the literature. 
This general-relativistic window function is essential for the correct calculation of the $\mu T$ cross-correlation in Section \ref{sec:leading_effect}. 
A series of technical details are collected in \mbox{Appendix \ref{app:appendix-1-A}.}


\subsection{A master equation for the production of \texorpdfstring{$\mu$}{\textbackslash mu} distortion}
\label{ssec:master_equation}

\noindent Following our assumptions, we model the electron-photon-baryon plasma as a single, viscous fluid with four velocity
$U^\mu$ and temperature $T(x)$. After $z_{i}\simeq 2\times 10^{6} $, 
all processes that change photon number become inefficient and the number of photons $ n $ 
leads to a covariantly conserved local current $ \nabla_{\mu} N^{\mu}=0 $.\footnote{We omit the label 
$ \gamma $ for the photon number current because this is the only current we will be \mbox{interested in.} }
This conserved quantity is the statistical conjugate of the dimensionless chemical potential $\mu(x)$. 
Kinetic equilibrium ceases to be a good approximation when Compton scattering becomes in efficient at redistributing momentum, 
around $z = z_f\simeq5\times10^4$. 
After $z = z_f$, the released energy goes into different kinds of distortions 
(the so-called $r$ distortions and $y$ distortions: we refer to \cite{Chluba:2011hw,Khatri:2012tw,Khatri:2013dha} and references therein for details). 
The transitions around $ z_{i} $ and $ z_{f} $ are actually smooth and can be captured quantitatively using the Green's functions fits of \cite{Chluba:2013vsa}. 
In the following, we derive a formula for the evolution of $\mu$ during this period. 

Viscous corrections lead to the dissipation of acoustic waves in the plasma and generate entropy. 
Such entropy increase cannot be balanced by a change in the number of photons, 
so the average energy and entropy per photon grow. 
Working at linear order in $\mu$, we can relate the chemical potential to the specific entropy (the entropy per particle), 
namely the ratio between the rest-frame entropy density $s$ and number density of photons $n$:
\begin{equation}
\label{eq:mu_from_s_and_n}
\frac{s}{n} = \frac{2\pi^4}{45\zeta(3)} \left[ 1+(A_n-A_s)\mu \right]\,\,,
\end{equation}
where we define $A_s\equiv \frac{135\zeta(3)}{2\pi^4}$, $A_n\equiv \frac{\pi^2}{6\zeta(3)}$ following \cite{Pajer:2013oca}. 
This equation tells us that we can compute the evolution of $\mu$ if we know how $s$ and $n$ evolve. 
For a perfect fluid, we know that the ratio $s/n$ will be constant along the fluid lines ($ n $ is conserved during the $\mu$-era). 
In presence of viscous corrections, however, the conservation of entropy and photon number density take the form 
\cite{Weinberg:1971mx,Weinberg:2008zzc,Pajer:2013oca}\footnote{The original derivation 
in \cite{Weinberg:1971mx} used the convention in which $ U^{\mu} $ is the velocity of particle transport, 
in which case $ \Delta N^{\mu}=0 $. Instead, in \cite{Weinberg:2008zzc,Pajer:2013oca} the velocity refers to the transport of energy, 
defined by the condition $ U^{\mu}\Delta T_{\mu\nu}=0 $. In this case $ \Delta N^{\mu}\neq 0 $. We use here the latter convention.} 
\begin{subequations}
\label{eq:entropy_and_number_conservation_with_viscosity}
\begin{align}
&\nabla_\mu(nU^\mu + \Delta N^\mu) = 0\,\,, \label{eq:entropy_and_number_conservation_with_viscosity-1} \\
&\nabla_\nu(s U^\nu + \mu\Delta N^\nu) = -\frac{\Delta T^{\mu\nu}\nabla_\nu U_\mu}{T} + 
\Delta N^\nu\nabla_\nu\mu\,\,, \label{eq:entropy_and_number_conservation_with_viscosity-2}
\end{align}
\end{subequations}
where $\Delta N^\mu$ and $\Delta T^{\mu\nu}$ are the leading 
(in an expansion in $t_\gamma\times\partial_\mu$, 
with $t_\gamma = (\sigma_\text{T}n_e)^{-1}$ being the photon mean free path) 
viscous corrections to the photon number density current and the stress-energy tensor. 
In accordance with our approximations, discussed in Section \ref{sec:general}, 
in writing \eqsI{entropy_and_number_conservation_with_viscosity} we have neglected the contribution of baryon (and electron) conservertion to the total entropy, 
which is indeed suppressed by the baryon-to-photon number ratio $ r $. Using \eqsI{entropy_and_number_conservation_with_viscosity}, 
we arrive at
\begin{equation}
\label{eq:derivative_of_s_over_n}
\begin{split}
&U^\mu\nabla_\mu\bigg(\frac{s}{n}\bigg) = -\frac{\mu}{n}\nabla_\nu\Delta N^\nu 
- \frac{\Delta T^{\mu\nu}\nabla_\nu U_\mu}{nT} + \frac{s}{n^2}\nabla_\mu\Delta N^\mu \\
&\hphantom{U^\mu\nabla_\mu\bigg(\frac{s}{n}\bigg) } = - \frac{\Delta T^{\mu\nu}\nabla_\nu U_\mu}{nT} 
+ \frac{\pi^6}{45\zeta(3)^2T^3}\nabla_\mu\Delta N^\mu + \mathcal{O}(\mu^2)\,\,,
\end{split}
\end{equation}
where we used the fact that $\Delta N^\mu = \mathcal{O}(\mu)$ and stopped at linear order in $\mu$. 
Indeed, as shown in \cite{Weinberg:1971mx,Weinberg:2008zzc,Pajer:2013oca}, 
$\Delta N^\mu$ and $\Delta T^{\mu\nu}$ take the form
\begin{subequations}
\label{eq:viscous_corrections}
\begin{align}
&\Delta N_\nu = -\chi\bigg(\frac{nT}{\rho + p}\bigg)^2P_\nu^{\hphantom{\nu}\rho}\nabla_\rho\mu\,\,, \label{eq:viscous_corrections-1} \\
&\Delta T_{\mu\nu} = -2\eta P^\rho_{\hphantom{\rho}(\mu}\nabla_{\rho} U_{\nu)} 
+ \bigg(\frac{2\eta}{3} - \zeta\bigg)\nabla_\rho U^\rho P_{\mu\nu}\,\,, \label{eq:viscous_corrections-2}
\end{align}
\end{subequations}
where the coefficients $\chi$, $\eta$ and $\zeta$ are, respectively, 
the heat conduction, shear viscosity and bulk viscosity, 
and $P^\mu_{\hphantom{\mu}\nu}\equiv\delta^\mu_{\hphantom{\mu}\nu} + U^\mu U_\nu$ is the projector on the instantaneous rest frame of the fluid. 
As shown in \cite{Pajer:2013oca}, $\Delta N^\mu$ vanishes up to terms of order of the baryon loading, 
defined as $R\equiv 3{\rho}_\text{b}/(4{\rho}_{\gamma})$. 
Additionally, the bulk viscosity is also suppressed by $r^2\simeq\num{d-19}$ with respect to the shear viscosity. 
For this reason, $\zeta$ can be safely neglected in the following \cite{Pajer:2013oca}. 
For the moment, we drop also the heat conduction $\chi$: its effect will be discussed in more detail in Section \ref{subsec:contributions_to_mu_T}. 
Therefore, with only the shear viscosity remaining, \eq{derivative_of_s_over_n} becomes
\begin{equation}
\label{eq:mu_generation-A}
U^\mu\nabla_\mu\bigg(\frac{s}{n}\bigg) = 
\frac{2\eta}{nT}\bigg({P^\rho_{\hphantom{\rho}(\mu}\nabla_{\rho} U_{\nu)}}\nabla^\nu U^\mu 
- \frac{(\nabla_\mu U^\mu)^2}{3}\bigg)\,\,,
\end{equation}
where $\eta$ is equal to $\frac{16}{45}t_\gamma\rho_\gamma$. 
Using the fact that $\nabla_\mu U_\nu = H P_{\mu\nu}$ on an FLRW background, 
it is straightforward to see that the right-hand side of the above equation starts at second order in perturbations. 
Moreover, \eq{mu_generation-A} explicitly shows that $\mu$ can be generated only if viscous corrections are present: 
solving for $\mu$ perturbatively in $t_\gamma\times\partial_\mu$ and stopping at first order in this expansion allows us to 
evaluate the thermodynamical quantities in the pre-factor on the right-hand side at zeroth order in $\mu$: 
\begin{equation}
\label{eq:mu_generation-B}
U^\nu\nabla_\nu\mu = \frac{8t_\gamma}{15(A_n - A_s)}\bigg({P^\rho_{\hphantom{\rho}(\mu}\nabla_{\rho} U_{\nu)}}\nabla^\nu U^\mu 
- \frac{(\nabla_\rho U^\rho)^2}{3}\bigg)\,\,.
\end{equation}
This expression, which is valid in the hydrodynamical approximation\footnote{In the perfect fluid limit, 
\ie $ t_{\gamma}=0 $, \eq{mu_generation-B} tells us that $\mu$ is simply frozen along the fluid lines, 
\ie $U^\rho\nabla_\rho\mu=0$. To finite order in $ t_{\gamma} $, the equation describes not only the generation of $ \mu $, 
but also the damping of $ \mu $ inhomogeneities due to viscosity \cite{Pajer:2013oca}, during the $ \mu $-era. 
The damping and evolution of $\mu$ inhomogeneities after the $ \mu $-era will be discussed in Section \ref{sec:after_mu_era} and Appendix \ref{sec:mu_damping}.} 
to all orders in perturbations, can be thought of as a generalization of Eq. (1.7) in \cite{Pajer:2013oca}. 
The generation of $\mu$ will proceed in time according to \eq{mu_generation-B}.


\subsection{\texorpdfstring{$\mu$}{\textbackslash mu} production at quadratic order}
\label{ssec:mu_quadratic}

\noindent\eq{mu_generation-B} can be solved at leading order in cosmological perturbation theory, 
to arrive at an expression for the generated $\mu$ distortion in terms of the primordial fluctuations. 
The final result is that $\mu = \mathcal{O}(\zeta^{2})$ can be written in terms of window function $ W $ as 
(we use the shorthand $\int_{\vec{k}}$ to denote $\int\frac{\dif^3k}{(2\pi)^3}$)
\begin{equation}
\label{eq:window_function-A}
\mu^{(2)}(\eta_f,\vec{x}) = \int_{\vec{k}_1}\int_{\vec{k}_2}\zeta(\vec{k}_1)\zeta(\vec{k}_2)W(\vec{k}_{1},\vec{k}_{2})e^{i(\vec{k}_1+\vec{k}_2)\cdot\vec{x}}\,\,,
\end{equation}
where the general-relativistic Fourier space window function\footnote{Although most of the dissipation takes place on sub-Hubble scales, 
it is very important for us to use the fully relativistic $ W $ given above in our calculation. 
In fact, the sub-Hubble approximation, \textit{e.g.} Eq. (6) in \cite{Pajer:2012vz} 
has the wrong scaling as one of the two wavenumber goes to zero. 
Using this sub-Hubble approximation instead of $ W $ gives a very small corrections in the computation of 
$ \braket{\mu T} $ from primordial non-Gaussianity. 
On the other hand, it is essential to use the full $ W $ when computing $\braket{\mu^{(2)}T^{(2)}}$, 
which is of one the second-order contributions to $ \mu T $. } (derived in Appendix \ref{app:appendix-1-A})
\begin{equation}
\label{eq:window_function-B}
\begin{split}
W(\vec{k}_{1},\vec{k}_{2})\equiv\frac{8}{15(A_{s}-A_{n})} &\left[ \left(\vec{k}_{1}\cdot \vec{k}_{2}\right)^{2}-\frac{1}{3}k_{1}^{2}k_{2}^{2} \right] \\
&\times \int_{0}^{\infty}\mathrm{d}z' \left(\frac{t_{\gamma}}{\mathcal{H}}\right)
\frac{T_{v}(z',k_{1})\,T_{v}(z',k_{2})}{\mathcal{H}^{2}} \mathcal{J}_{\mu}(z')\,\,.
\end{split}
\end{equation}
Here, $T_v$ is the transfer function of the velocity potential and $ \mathcal{J}_{\mu} $ is an analytic fit to the time window function 
computed numerically in \cite{Chluba:2013vsa}. 
Moreover, we recall that during radiation domination the damping scale and the photon mean free path $t_\gamma = (\sigma_\text{T}n_e)^{-1}$ 
are related by the approximate expression 
$k_\text{D}^{-1}\simeq\sqrt{t_\gamma\eta/a} = \sqrt{t_\gamma/a\mathcal{H}}$, 
\ie $k_\text{D}^2\simeq a\mathcal{H}/t_\gamma$ \cite{Dodelson:2003ft}. 

The window function of \eq{window_function-B} agrees with, for instance, Eq. (1.7) in \cite{Pajer:2013oca}, 
even though that was derived for sub-Hubble scales only. 
Notice that when the angle between the momenta in this window function is zero 
(which happens for instance upon taking an ensemble average of $\mu$ itself), 
this window function also matches the window function in \cite{Chluba:2016aln}: 
however, this reference \textit{does not} write the full window function including super-Hubble modes, which is essential to this work. 
Moreover, we stress that the spatial structure in our formula differs from \cite{Chluba:2016aln} for generic momenta. 
From \eq{window_function-B} we also see how this window function has the expected behavior when either one of the two wavenumbers goes to zero, 
namely $ W(k_{1},k_{2})\sim k_{1}^{2} $ for $k_{1}\ll \mathcal{H}$, 
and similarly for $ k_{2} $. In addition, sub-Hubble modes much longer than the Silk damping scale during the $\mu$-era, 
$ \cH \ll k_{1,2}\ll k_{\rm D} $ are suppressed by $k_1 k_2/k^2_{\text{D}}(z_f)$, 
where $k_{\rm D}$ at the end of the $\mu$-era is roughly $k_{\text{D}}(z_f)\simeq\mpc{50}$.


\subsection{\texorpdfstring{$\mu$}{\textbackslash mu} production at cubic order}
\label{ssec:mu_cubic}

\noindent We now estimate the production of $ \mu $ at cubic order, $ \mu^{(3)} $, 
which in Section \ref{sec:leading_effect} we will find to give the leading contribution to the $ \mu T $ correlation in single-field inflation. 
One can write the dependence of the expectation value of $\mu_{S}$ on a long $\zeta$ mode as 
\begin{equation}
\label{eq:bias_1}
\braket{\mu_{S}}_{\zeta_{L}}(\eta,\vec{x})=\braket{\mu_{S}}_{\zeta_{L}=0}-\frac{b_{1}}{\Lambda^2}{\partial^{2}}\zeta_{L}(\eta,\vec{x})
\braket{\mu_{S}}_{\zeta_{L}=0} + \text{higher derivatives}\,\,,
\end{equation} 
where $\partial^2=\delta_{ij}\partial_i\partial_j$ and 
\begin{equation}
\label{eq:bias_2}
\frac{b_{1}}{\Lambda^{2}}\equiv{-\frac{1}{\braket{\mu_{S}}_{\zeta_{L}=0}}} 
\bigg[\frac{\partial\!\braket{\mu_{S}}_{\zeta_{L}}}{\partial(\partial^{2}\zeta_{L})}\bigg]_{\zeta_{L}=0} \,\,.
\end{equation}
Here we have in mind $b_{1}$ of order unity (the sign has been chosen for later convenience: 
more precisely, with this choice we will have a contribution $C^{\mu T}_\ell\sim b_1$ in Section \ref{sec:forecast}). 
Notice that the physical situation here is the same as the one that leads to the bias expansion in the 
context of galaxy clustering (see \cite{Desjacques:2016bnm} for a comprehensive review), 
\ie that of a separation between the scale of local physics and the scale at which we measure correlation functions: 
for this reason, in the following we will refer to $b_1$ as the ``bias'' parameter. 

The question now is: given that we want $ b_{1}\sim\mathcal{O}(1) $, 
what is the value of the scale $ \Lambda $ that suppresses $\partial^2 \zeta_{L}$? 
To answer this, first observe that spatial curvature of the local, ``separate'' universe is related to the second derivative of $\zeta$ as \cite{Baldauf:2011bh,Dai:2015jaa}
\begin{equation}
\frac{\partial^{2}\zeta}{\mathcal{H}^{2}} \sim \Omega_{K}\,\,.
\end{equation}
Then, since spatial curvature modifies the evolution of the Hubble rate and the evolution of short scale modes at order unity, 
we expect the suppression scale $\Lambda$ to be approximately the (comoving) Hubble scale at the end of the $\mu$-era (where it is smallest), 
which we call $\mathcal{H}_{f}$. Alternatively, one could argue that at leading order in $t_{\gamma}$, 
there is no additional scale available besides Hubble. 
This is seen, for example, from the general-relativistic expression for the window function: 
if both modes are sub-Hubble, \ie $\mathcal{H}\ll k_{1,2} \ll k_{\rm D}$, 
the suppression is $k_{1}k_{2}/k_{\rm D}^{2}$, manifestly showing that $\mu$ is created by damping. 
If both modes are super-Hubble, it becomes $k_{1}^{2}k_{2}^{2}/k_{\rm D}^{2}\mathcal{H}^{2}$. 
We here care about the third-order contribution, with at least one super-Hubble mode. 
The dominant contribution then comes from the other two modes being sub-Hubble. 
Its scaling is $k_{3}^{2}/\mathcal{H}^{2}\times k_{1}k_{2}/k_{\rm D}^{2}$, 
which leads to the above suppression when the short modes are of order $k_{\rm D}$.

\section{Evolution of \texorpdfstring{$\mu$}{\textbackslash mu}-type spectral distortion}
\label{sec:after_mu_era}

\noindent We are now ready to discuss the evolution of the chemical potential after the end of the $\mu$-era, 
and how it contributes to the $\mu T$ correlator. 
As we have seen in the previous section, 
$\mu$ is sourced by perturbations close to the damping scale, 
which is much shorter than the Hubble radius. We will denote these short scales with a subscript $S$. 
Consider then a long wavelength $ \zeta_{L} $ mode, denoted by a subscript $L$, 
well outside the Hubble radius at the time $\eta_f$ of the end of the $\mu$-era, but inside the Hubble radius today. 
We know that its effect on short-scale perturbations sourcing $\mu$, 
at zeroth and linear order in the gradients of this long mode, 
will be equivalent to that of a coordinate transformation 
(see, \textit{e.g.}, \cite{Maldacena:2002vr,Creminelli:2004yq}). 
More precisely, we start by focusing on the zeroth order in gradients. 
We will discuss the general case at the end of the section.


\subsection{From CFC to global coordinates}
\label{ssec:after_mu_era}

\noindent As we discussed in Section \ref{ss:generalstrategy}, the calculation of short-long mode coupling during the $ \mu $-era 
is most easily performed in local (CFC) coordinates. 
On the other hand, eventually the long mode re-enter the Hubble radius and induces the modulation in $ T $ and $ \mu $ which we aim to measure. 
Therefore, at some point before observation, we have to change from CFC to global coordinates 
to account for how the long mode affect the propagation of the photons we eventually observe. 
We choose to do so at the end of the $ \mu $-era, corresponding to conformal time $ \eta=\eta_{f} $. Using the fact that we are deeply into radiation dominance, 
the coordinate change from CFC to global coordinates reads as \cite{Weinberg:2003sw,Creminelli:2004pv,Creminelli:2011sq,Mirbabayi:2014hda} 
\begin{subequations}
\label{eq:adiabatic_mode_in_radiation_domination}
\begin{align}
&\tilde{\eta} = \bigg(1-\frac{\zeta_L}{3}\bigg)\eta\,\,, \label{eq:adiabatic_mode_in_radiation_domination-1} \\
&\tilde{\vec{x}} = (1+\zeta_L)\vec{x}\,\,, \label{eq:adiabatic_mode_in_radiation_domination-2}
\end{align}
\end{subequations}
where the long mode $\zeta_L$ is absent in the $\tilde{x}$ coordinates.\footnote{We note that these formulas for the so-called adiabatic mode 
have been derived considering the universe as composed of a single perfect fluid, \ie neglecting viscosity, which is indeed negligible on very large scales.} 
After this transformation, the metric takes the form 
\begin{equation}
\label{eq:poisson_metric}
\dif s^2 = a^2(\eta)\big[{-e^{2\Phi}\dif\eta^2} + e^{-2\Psi}\dif{\vec{x}}^2\big]\,\,,
\end{equation}
with
\begin{subequations}
\label{eq:potentials}
\begin{align}
&\Phi = \Phi_S - \frac{2\zeta_L}{3} - \frac{\eta\zeta_L}{3}\frac{\partial\Phi_S}{\partial\eta} 
+ \zeta_Lx^i\frac{\partial\Phi_S}{\partial x^i}\,\,, \label{eq:potentials-1} \\
&\Psi = \Psi_S - \frac{2\zeta_L}{3} - \frac{\eta\zeta_L}{3}\frac{\partial\Psi_S}{\partial\eta} 
+ \zeta_Lx^i\frac{\partial\Psi_S}{\partial x^i}\,\,. \label{eq:potentials-2}
\end{align}
\end{subequations}
\ie we recognize the Poisson gauge with no vector or tensor modes. 
Under the change of coordinates of \eqsI{adiabatic_mode_in_radiation_domination}, 
$\mu_S$ transforms as a scalar. 
More precisely, we have
\begin{equation}
\label{eq:mu_transformation}
\mu = \mu_S - \frac{\eta\zeta_L}{3}\frac{\partial\mu_S}{\partial\eta} + \zeta_Lx^i\frac{\partial\mu_S}{\partial x^i}\,\,.
\end{equation}
Given that the conservation of $\mu_S$ after $\eta_f$ implies $\frac{\partial\mu_S}{\partial\eta} = 0$, 
we will drop the time derivative in \eq{mu_transformation}. 
The fluid velocity $U^\mu$, instead, takes the form \cite{Creminelli:2011sq}
\begin{equation}
\label{eq:fluid_velocity}
v^i = v^i_S - \frac{\eta\zeta_L}{3}\frac{\partial v^i_S}{\partial\eta} + \zeta_Lx^j\frac{\partial v^i_S}{\partial x^j}\,\,,
\end{equation}
where $U^i = e^{\Psi} v^i/a$. 
We see that, since we stop at zeroth order in the gradients of $\zeta_L$, 
no large-scale velocity is generated.

\eqsIII{potentials}{mu_transformation}{fluid_velocity} 
give us the initial condition for the evolution of $\mu(\eta_f,\vec{x}_f)$ on the hypersurface $\eta=\eta_f$, 
when $\mu$ stops being created, up to the observer at $(\eta_0,\vec{x}_0)$.


\subsection{Non-linear evolution}
\label{ssec:nonlinear_evolution}

\noindent Let us now assume that the evolution from the end of the $\mu$-era 
up to the last-scattering surface is dictated by the conservation of $\mu$ along the fluid lines 
(we generalize this in Appendix \ref{sec:mu_damping}). That is, until recombination the chemical potential satisfies the equation
\begin{equation}
\label{eq:fluid_lines_conservation}
U^\rho\nabla_\rho\mu = 0\,\,.
\end{equation}
After decoupling, the evolution is dictated only by the conservation of $\mu$ along the photon geodesics. 
Indeed, while the CMB temperature at recombination experiences an additional redshift $E_0/E_\text{rec}$ 
as the photons travel from the last scattering surface to us 
(which is the source of the Sachs-Wolfe, integrated Sachs-Wolfe and Doppler effects), 
$\mu$ is immune to this effect. 
In fact, the solution to the collisionless Boltzmann equation $\frac{\text{D}f}{\dif\lambda} = 0$ 
(where $\lambda$ is an affine parameter along the photon geodesics) 
is given by 
\begin{equation}
\label{eq:collisionless_boltzmann}
\frac{2}{e^{\frac{E_\text{rec}}{T_\text{rec}} + \mu_\text{rec}}-1} = 
\frac{2}{e^{\frac{E_0}{T_0} + \mu_0}-1}\,\,,
\end{equation}
where we used the Bose-Einstein expression for the photon distribution function, with 
$E$ being the photon energy for an observer at rest with respect to the CMB frame 
(who coincides with the an observer comoving with the fluid during the tight-coupling regime). 
Since the equality must hold for all values of the photon energy, we obtain (for the temperature-only case, 
see also Appendix A of both \cite{Creminelli:2011sq} and \cite{Mirbabayi:2014hda})
\begin{equation}
\label{eq:solution_to_collisionless_boltzmann}
T_0 = \frac{E_0}{E_\text{rec}}T_\text{rec}\,\,,\qquad\mu_0 = \mu_\text{rec}\,\,.
\end{equation}
We stress that this relation holds to all orders in cosmological perturbations, since it is just a consequence of Liouville's theorem. 

The presence of the long mode will perturb the fluid lines and the photon trajectories, 
and with them the expression of the spatial coordinates on the $\eta_f$ hypersurface in terms of the coordinates at the observation point:
\begin{itemize}
\item At linear order in the long mode, the solution for the fluid lines would take the form $x^i(t) = x^i + \int_{t_f}^t\dif s\,v^i_L(s,\vec{x})$. 
However, as we see in \eq{fluid_velocity}, $v^i_L$ vanishes at zeroth order in gradients, so that $\vec{x}_f = \vec{x}_\text{rec}$. 
\item After decoupling, the solution of the geodesic equation at linear order in the long mode gives \cite{Creminelli:2011sq}
\begin{equation}
\label{eq:x_rec_vs_x_obs}
\begin{split}
&\vec{x}_\text{rec}(\vers{n}) = \vec{x}_0 + \vers{n}(\eta_0-\eta_\text{rec}) 
+ 2\vers{n}\int_{\eta_\text{rec}}^{\eta_0}\dif\eta\,\Phi_L(\eta,\vec{x}_\eta) 
- 2\int_{\eta_\text{rec}}^{\eta_0}\dif\eta\,(\eta-\eta_\text{rec})\vec{\nabla}\!_\perp\Phi(\eta,\vec{x}_\eta)\,\,,
\end{split}
\end{equation}
where $\vers{n}$ is the direction of observation, 
the time integrals are along the unperturbed photon trajectory $\vec{x}_\eta=\vers{n}(\eta_0-\eta)$, 
and $\vec{\nabla}\!_\perp$ denotes a derivative perpendicular to the line of sight, 
$\nabla^\perp_i = (\delta_{ij} - \hat{n}_i\hat{n}_j)\partial_j$. 
Since $\vec{x}_f = \vec{x}_\text{rec}$, \eq{x_rec_vs_x_obs} provides the relation between $\vec{x}_0$ and $\vec{x}_f$ and, 
with it, the relation between the inhomogeneities in $\mu$ at the end of the $\mu$-era to the anisotropies in $\mu$ seen in the sky today. 
\end{itemize}

To summarize, combining \eq{x_rec_vs_x_obs} with \eq{mu_transformation} 
we obtain the full solution for $\mu(t_0,x_0)$ at the observer's point, \ie
\begin{equation}
\label{eq:mu_observed}
\mu(\eta_0,x_0) = \mu_S(\eta_f,\vec{x}_\text{rec}) - \frac{3}{2}\Phi_L(\eta_f,\vec{x}_\text{rec})\,\vers{n}\cdot\vec{\nabla}\!_{\vers{n}}\,\mu_S(\eta_f,\vec{x}_\text{rec})\,\,,
\end{equation}
where we used the zeroth-order geodesic equation, $\vec{x}_\text{rec} = \vers{n}(\eta_0 - \eta_\text{rec})$, 
to rewrite $\vec{x}_f\cdot\vec{\nabla}\!_{\vec{x}_f}$ in \eq{mu_transformation}, 
and we used the fact that in radiation dominance on super-Hubble scales $\zeta_L = -3\Phi_L/2$. 
Moreover, if we focus on large angular scales $\ell\lesssim 100$, 
we can rewrite the long-wavelength Newtonian potential as $-9\Theta_L(\vers{n})/2$, 
using the Sachs-Wolfe approximation. 

The above solution for the observed $\mu$ anisotropies straightforwardly shows that the angular correlator $C^{\mu T}_\ell$ vanishes. 
Indeed, we have seen that the only effects of the long mode on the evolution after $\eta_f$ 
is the modification of the relation between $\vec{x}_0$ and $\vec{x}_f$ and the spatial derivative term coming from 
the effect of the long mode on the short modes at $\eta_f$, 
\ie the second term on the right-hand side of \eq{mu_observed}. 
Consider then the ensemble average of $\braket{\mu_S}$ at $\eta_f$: 
since it does not depend on spatial coordinates, it is unaffected by \eqsII{x_rec_vs_x_obs}{mu_observed}. 
The expression for $\braket{\mu T}$, then, schematically reads 
\begin{equation}
\label{eq:schematic_correlator}
\braket{\mu T}\sim\braket{\Theta_L}\braket{\mu_S} + \braket{\Theta_L^{2}}\vec{\nabla}\!_{\vers{n}}\braket{\mu_S} = \braket{\Theta_L}\braket{\mu_S} = 0\,\,,
\end{equation}
since $\braket{\mu_S}$ does not contain any long mode for $\Theta_L$ to correlate with. 
A more detailed proof of \eq{schematic_correlator} is given in Appendix \ref{app:appendix-2}. 

As in \cite{Creminelli:2011sq}, the result of \eq{mu_observed} is valid if the long mode is 
outside the sound horizon at the end of the $\mu$-era (but inside the Hubble radius today), 
and holds at zeroth order in an expansion in $k_L/k_S$, 
where $k_L\sim\ell/\eta_0$ is the long-wavelength temperature mode and 
$k_S\sim k_\text{D}(z_f)\simeq\mpc{50}$ is the damping scale at the end of the $\mu$-era. 
We can now see that there are two very important differences with respect to the computation of squeezed CMB bispectrum 
$B_{\ell_L\ell_S\ell_S}$ of \cite{Creminelli:2011sq}: 
\begin{itemize}
\item While the expression for $B_{\ell_L\ell_S\ell_S}$ of \cite{Creminelli:2011sq} holds only in the squeezed limit, 
\eq{mu_observed} (and consequently \eq{schematic_correlator}) is an expression for the \textit{full} $\mu T$ angular correlator.
\item In the $B_{\ell_L\ell_S\ell_S}$ case the long mode needed to be outside the Hubble radius at recombination, 
limiting the validity of the calculation to $\ell_L\lesssim{100}$. 
Here $\ell$ can be pushed up to $\ell\sim\cH_f\eta_0 = \mathcal{O}({1500})$, 
since it is enough to consider the long mode to be outside the Hubble radius at the end of the $\mu$-era.
\end{itemize}


\subsection{First order in the gradient of the long mode and beyond}
\label{ssec:quadratic_order_and_beyond}

\noindent In this section we show that our results can be directly extended to include gradients of the long mode. 
While to derive the explicit solution of \eq{mu_observed} we stopped at zeroth order in $k_L/k_S$, 
the equations \eqsII{fluid_lines_conservation}{solution_to_collisionless_boltzmann} are non-perturbative in such expansion. 
The solutions to these equations, 
\ie conservation of $\mu$ along the fluid lines up to the last-scattering surface, and along the photon geodesics up to the observer's point, 
always involve spatial derivatives of $\mu$: 
once projected on the sky, then, these spatial derivatives turn into derivatives along the direction of observation $\vers{n}$,\footnote{We can see this by 
using the solution of \eq{x_rec_vs_x_obs}, \ie $\vec{x}_\text{rec}(\vers{n}) = \vec{x}_0 + \vers{n}(\eta_0-\eta_\text{rec})$ on the background.} 
so that they will vanish once we average over the short modes. 
This can be seen as a generalization of the intuitive argument put forward in Section \ref{sec:introduction}. 
Therefore, we are limited only by how accurately Weinberg's theorem describes the evolution of the chemical potential up to the end of the $\mu$-era. 
As shown in \cite{Creminelli:2012ed,Hinterbichler:2012nm,Creminelli:2013mca} (see also \cite{Mirbabayi:2014hda}), 
Weinberg's theorem can be extended to gradient order: 
\textit{i.e} also at this order the effect of the long mode on the short modes is equivalent to a coordinate transformation and, 
similarly to the discussion below \eq{mu_transformation}, $\zeta_L$ couples only to spatial derivatives of $\mu_S$. 
At order $k_L^2/\cH_f^2$, however, we cannot anymore use this argument: 
indeed, at this order the long mode is contributing to the local curvature, 
and this effect cannot be mimicked by a coordinate transformation. 
This is precisely the CFC result of Section \ref{ssec:mu_cubic}. 

Throughout this section we have assumed that after the end of the $\mu$-era 
the evolution of the chemical potential is simply dictated by \eq{fluid_lines_conservation}. 
However, as shown in \cite{Pajer:2013oca}, 
during the tight-coupling evolution up to the last-scattering surface inhomogeneities of $\mu$ on small scales will be damped. 
In \mbox{Appendix \ref{sec:mu_damping}}, we show that these effects do not give additional correlations with the long mode.

\section{Observed \texorpdfstring{$\mu T$}{\textbackslash mu T} cross-correlation}
\label{sec:leading_effect}

\noindent We have argued that the observed $\mu T$ correlator 
receives no ``projection'' contributions from modes outside the Hubble radius during the $\mu$-era up 
to corrections of order $ k_{L}^{2}/\cH_{f}^{2}\ll 1 $. 
In particular, these non-primordial effects have a different $ k $ dependence 
(and therefore $\ell$ dependence in angular correlators) from local non-Gaussianity. 
The above results were derived assuming the only source of $\mu$ production is Silk damping. 
In this section, we estimate also the production due to heat conduction and compare all contributions 
(like, \textit{e.g.}, those from temperature non-linearities). 
After providing formulae for the observed $\mu T$ angular correlator we show, by mean of a Fisher forecast, 
that the non-primordial contributions do not need to be computed because marginalizing over them does not appreciably 
change the constraints of $ f_{\rm NL} $ even for a very futuristic cosmic variance-limited experiment.

\subsection{Contributions to the observed \texorpdfstring{$\braket{\mu T}$}{<\textbackslash mu T>}}
\label{subsec:contributions_to_mu_T}

\noindent We consider three types of non-primordial corrections, namely 
$\langle\mu^{(3)}T^{(1)}\rangle$, $\langle\mu^{(2)}T^{(2)}\rangle$, and $\langle\mu^{(1)}T^{(1)}\rangle$. 
In the following, we discuss these remaining sources separately, 
estimating their size and finding which one gives the strongest contribution. 
All of these effects must then be compared to the signal we are interested in, 
\ie local non-Gaussianity from multi-field inflation (or any model that violates the consistency relation). 
For simplicity, we estimate and compare all contributions to the $\braket{\mu T}$ correlator evaluated at the last-scattering surface, $\eta=\eta_{\rm rec}$. 
We find that the largest non-primordial contribution to $ \mu T $, 
\ie the leading contribution in a single-field universe obeying the consistency relation, 
comes from $\mu$ production at third order, discussed in Section \ref{ssec:mu_cubic}.

\subsubsection*{Local non-Gaussianity}

\noindent We find it convenient to discuss the size of any non-primordial contribution to $ \mu T $ in terms of an \textit{effective} $ f_{\rm NL} $. 
Therefore here we briefly review the contribution of (multi-field) local non-Gaussianity to $ \mu T $, defined as\footnote{Notice that, 
since we are working in CFC, $\fnl$ equals zero in single-field cosmologies.}
\begin{equation}
\label{eq:local_bispectrum}
\braket{\zeta(\vec{k}_1)\zeta(\vec{k}_2)\zeta(\vec{k}_3}'={\frac{6\fnl}{5}}\big[P_\zeta(k_1)P_\zeta(k_2)+\text{$2$ perms.}\big]\,\,,
\end{equation}
where we use a prime to denote that we have factored out a delta function $(2\pi)^3\delta^{(3)}(\sum_i\vec{k}_i)$ of momentum conservation.
Since $\mu_{S}$ starts quadratic in $\zeta$, at first order in $\fnl$ the bispectrum of \eq{local_bispectrum} leads to a contribution to 
$\langle\mu T\rangle$ of the form (c.f. Eq.~(16) in \cite{Pajer:2012vz}, and the computation in Appendix \ref{app:forecast})
\begin{equation}
\begin{split}
\langle\mu(\vec{x})T(\vec{y})\rangle&= \int_{\vec{q}_1,\,\vec{q}_2,\,\vec{k}} 
\langle\mu_{S}(\vec{q}_{1}-\vec{k},\vec{k})T(\vec{q}_2)\rangle e^{i(\vec{q}_1\cdot\vec{x}+\vec{q}_2\cdot\vec{y})} \\
&\sim \fnl\braket{\mu_{S}} \int_{\vec{q}}P_{\zeta}(q)\Delta^{T}(q)e^{i\vec{q}\cdot(\vec{x}-\vec{y})}\,\,,
\end{split}
\end{equation} 
where $\Delta^T(q)$ is the transfer function that relates $\zeta$ to temperature fluctuations at the time of last-scattering 
(we neglect the equivalent for $\mu$, responsible for the damping of $\mu$ fluctuations, since it does not affect our estimates). 
Its Fourier transform then reads
\begin{equation}
\label{eq:fnl_naive}
\langle\mu(\vec{q})T(-\vec{q})\rangle^{\prime}_{\text{NG}}\sim \fnl\braket{\mu_{S}}P_\zeta(q) \Delta^{T}(q)\,\,.
\end{equation}

\subsubsection*{\texorpdfstring{$\mu$}{\textbackslash mu} production at third order}
\label{mu3}

\noindent As outlined in Section \ref{sec:after_mu_era}, 
any mode coupling that is induced between long modes and $\mu_{S}$ after the $\mu$-era drops out of the $\mu T$ correlator. 
The leading contribution therefore comes from the second derivative of the long mode during the $\mu$-era, 
which is locally equivalent to a spatial curvature and modulates the amount of distortion that is produced. 
The relevant cubic term $ \mu^{(3)} $ was estimated in Section \ref{ssec:mu_cubic} to be
\begin{equation}
\mu^{(3)}(\eta,\vec{x})=-b_{1} \,\frac{\partial^{2}\zeta_{L}(\eta,\vec{x})}{\cH^{2}_f} \braket{\mu_{S}}\,\,.
\end{equation}
Its contribution to the correlator is then given by
\begin{equation}
\label{eq:mu_three}
\langle\mu^{(3)}(\vec{q})T^{(1)}(-\vec{q})\rangle^{\prime}= b_{1}\left(\frac{q}{\mathcal{H}_{f}}\right)^{2}P_\zeta(q) \Delta^{T}(q)\braket{\mu_{S}}+\mathcal{O}(q^{3})\,\,.
\end{equation}

\subsubsection*{Second-order temperature corrections}

\noindent Here we provide a rough estimate of second order effects in temperature $ T^{(2)} $. 
Given that these are found to be very small corrections, we are cavalier about the second-order transfer function and simply use the rough approximation 
$T=T^{(1)} + ( T^{(1)} )^{2}$. This contributes as 
\begin{equation}
\begin{split}
\langle\mu^{(2)}(\vec{x})T^{(2)}(\vec{y})\rangle&=\int_{\vec{q}_1,\,\vec{q}_2,\,\vec{k}_{1},\,\vec{k}_{2}}
\langle W(\vec{q}_1-\vec{k}_{1},\vec{k}_{1})\zeta(\vec{q}_1-\vec{k}_{1})\zeta(\vec{k}_{1})T(\vec{q}_2-\vec{k}_{1})T(\vec{k}_{2})\rangle \\
&\hphantom{=\int_{\vec{q}_1,\vec{q}_2,\vec{k}_{1},\vec{k}_{2}}}\times e^{i(\vec{q}_1\cdot\vec{x}+\vec{q}_2\cdot\vec{y})} \\
&=\int_{\vec{q}}\int_{\vec{k}}W(\abs{\vec{q}-\vec{k}},k)P_{\zeta}(\abs{\vec{q}-\vec{k}})\Delta^{T}(|\vec{q}-\vec{k}|)e^{i\vec{q}\cdot(\vec{x}-\vec{y})}P_{\zeta}(k)\Delta^{T}(k)\,\,,
\end{split} 
\end{equation} 
where $W$ indicates the window function of Eq. \eqref{eq:window_function-B}, 
which contains a two derivative suppression factor when at least one of the arguments is small. 
Fourier transforming, we thus find 
\begin{equation}
\langle\mu^{(2)}(\vec{q})T^{(2)}(-\vec{q})\rangle^{\prime}=
\int_{\vec{k}} W(\abs{\vec{q}-\vec{k}},k) P_\zeta(\abs{\vec{q}-\vec{k}})\Delta^{T}(|\vec{q}-\vec{k}|)P_\zeta(k)\Delta^{T}(k)\,\,.
\end{equation}
We now show that this is subdominant with respect to $ \braket{ \mu^{(3)}T^{(1)}} $. 
Let us separate the integral into modes longer than the suppression scale during $\mu$ and those of order 
$q_\text{eff}\equiv\sqrt{\mathcal{H}_{f}k_{\mathrm{D}}}$, 
which is roughly the scale at which the window function qualitatively changes behavior. 
Then, dropping some $\mathcal{O}(1)$ numbers in the window function, we find
\begin{equation}
\label{eq:long_contribution}
\begin{split}
\langle\mu^{(2)}(\vec{q})T^{(2)}(-\vec{q})\rangle^{\prime}_\text{long}&\sim\int_{\vec{k},\,k\,\ll\,q_{\text{eff}}}P_\zeta(\abs{\vec{q}-\vec{k}})\Delta^{T}(|\vec{q}-\vec{k}|)P_\zeta(k)\Delta^{T}(k)
\left(\frac{\abs{\vec{q}-\vec{k}}}{q_\text{eff}}\right)^{2}\left(\frac{k}{q_\text{eff}}\right)^{2} \\
&\sim\Delta^{2}_{\zeta}(q)P_\zeta(q)\big(\Delta^{T}(q)\big)^2\left(\frac{q}{q_\text{eff}}\right)^{4}\,\,,
\end{split}
\end{equation} 
where we used $k\sim q$. The remaining part takes the form 
\begin{equation}
\begin{split}
\langle\mu^{(2)}(\vec{q})T^{(2)}(-\vec{q})\rangle^{\prime}_\text{short}&\sim
\int_{\vec{k},\,k\,\sim\,q_\text{eff}}\Delta^{T}(|\vec{q}-\vec{k}|)P_\zeta(\abs{\vec{q}-\vec{k}})\Delta^{T}(k)P_\zeta(k) \\
&\sim\Delta^{2}_{\zeta}(q_\text{eff})P_\zeta(q_\text{eff})\big(\Delta^{T}(q_\text{eff})\big)^{2}\,\,.
\end{split}
\end{equation} 
From this we can easily see that the long contribution of \eq{long_contribution} is subdominant to the above contribution. 
Comparing the short contribution to $\mu$ production at third order we find
\begin{equation}
\frac{\langle\mu^{(2)}(\vec{q})T^{(2)}(-\vec{q})\rangle^{\prime}}{\langle\mu^{(3)}(\vec{q})T^{(1)}(-\vec{q})\rangle^{\prime}}\sim
\left(\frac{q}{q_\text{eff}}\right)^{-2}\left(\frac{\mathcal{H}_{f}}{q_\text{eff}}\right)^{2}\frac{P_\zeta(q_\text{eff})}{P_\zeta(q)}\frac{\Delta^{T}(q_\text{eff})}{\Delta^{T}(q)}\lesssim
\left(\frac{q}{q_\text{eff}}\right)\left(\frac{\mathcal{H}_{f}}{q_\text{eff}}\right)^{2} \ll 1\,\,,
\end{equation}
where we use that $\ex{\mu_{S}}\sim \Delta_{\zeta}^{2}(q_{\text{D}})$.

\subsubsection*{\texorpdfstring{$\mu$}{\textbackslash mu} production at first order due to heat conduction}

\noindent If we look at \eqsII{derivative_of_s_over_n}{viscous_corrections-1}, we can 
see that the creation of $\mu$ at linear order due to heat conduction is suppressed by two spatial derivatives. 
Indeed, we see that the contribution to the evolution of $\mu$ from heat conduction is a divergence of a vector orthogonal to the fluid lines, 
so that up to first order in perturbations it will be $\sim\partial^2\zeta$ 
(see also Eq.~(3.18) and Eq.~(3.19) of \cite{Pajer:2013oca}). 
The suppression scale for this effect is $k_\mathrm{D}^{-2}$: the heat conduction coefficient is 
also proportional to the photon mean free path $t_\gamma$ \cite{Pajer:2013oca}. 
We therefore estimate its contribution from super-Hubble modes to be at most 
\begin{equation}
\label{esti}
\mu^{(1)}(\vec{q})\sim\left(\frac{q}{k_\mathrm{D}}\right)^{2}R^{2}\zeta(\vec{q})\,\,,
\end{equation}
where the baryon loading is $R = 3\bar{\rho}_\text{b}/(4\bar{\rho}_{\gamma})$. Inside the correlator this yields
\begin{equation}
\label{eq:baryon_term}
\langle\mu^{(1)}(\vec{q})T^{(1)}(-\vec{q})\rangle^{\prime}\sim \left(\frac{q}{k_\mathrm{D}}\right)^{2}R^{2}P_\zeta(q)\Delta^{T}(q)\,\,.
\end{equation}
Comparing this to the first contribution, we find
\begin{equation}
\frac{\langle \mu^{(1)}(\vec{q})T^{(1)}(-\vec{q})\rangle^{\prime}}{\langle \mu^{(3)}(\vec{q})T^{(1)}(-\vec{q})\rangle^{\prime}}\sim
\frac{R^{2}}{\Delta^{2}_{\zeta}(k_\mathrm{D})}\left(\frac{\mathcal{H}_{f}}{k_\mathrm{D}}\right)^{2}\,\,.
\end{equation}
Now note that $R$ at the end of the $ \mu $-era is approximately given by\footnote{Note that at earlier times, 
the comoving damping scale $k_\text{D}$ increases faster that the conformal Hubble rate, 
which means that the baryon term is smaller at earlier times during the $\mu$-era.}
\begin{equation}
R= R_{\rm eq}\frac{(1+z_\text{eq})}{(1+z_{f})}\simeq \frac{1}{6}\frac{3\times10^{3}}{5\times10^{4}}\simeq 10^{-2}\,\,,
\end{equation}
where again we have used that $\ex{\mu_{S}}\sim \Delta_{\zeta}^{2}(q_{\text{D}})$. 
Since $\Delta_{\zeta}^{2}\sim 10^{-9}$, the baryon suppression is clearly less than the additional perturbations, 
which forces us to compare the Hubble radius during the $\mu$-era and the damping scale as well. 
At the end of the $\mu$-era, 
$k_\mathrm{D}\simeq\mpc{50}$ and Hubble rate is approximately $\mpc{d-1}$. Hence the ratio is of order
\begin{equation}
\left(\frac{\mathcal{H}_{f}}{k_\mathrm{D}}\right)\simeq 10^{-3}\,\,,
\end{equation}
which means that the first order production is estimated to be \SI{10}{\percent} of the cubic production. 
Notice that this number is not very small and quite a few assumptions went into this estimate, 
such as $ b_{1} = \mathcal{O}(1) $ and Eq.~\eqref{esti}.

\subsection{\texorpdfstring{$\mu T$}{\textbackslash mu T} angular correlation and Fisher forecast}
\label{sec:forecast}

\noindent We have concluded that the largest contribution from non-linear evolution is the bias effect of \eqsII{bias_1}{bias_2}. 
Let us now first translate our result into angular correlations and then compare it with the contribution coming from local non-Gaussianity. 
More precisely, we carry out a Fisher forecast on the detectability of $\fnl$ by a full-sky cosmic variance-limited experiment that can measure $\mu$ 
anisotropies up to $\ell_{\rm max} = 1000$ (a similar forecast for a noise-dominated PIXIE-like experiment is carried out in Appendix \ref{app:PIXIE}). 

In order to write down the likelihood, we need the expression for the decomposition of $\mu$ on the sky in spherical harmonics, \ie
\begin{equation}
\label{eq:a_mu_ell_m}
a^\mu_{\ell m}(\eta_0,\vec{x}) = 4\pi\, i^{-\ell}\int_{\vec{k}}e^{i\vec{k}\cdot\vec{x}}\mu(\eta_f,\vec{k})\underbrace{e^{-k^2\Delta\qD^{-2}}j_\ell(k\Delta\eta)}_
{\hphantom{\Delta^\mu_\ell(k)\,}\equiv\,\Delta^\mu_\ell(k)}Y^\ast_{\ell m}(\vers{k})\,\,,
\end{equation}
where $\Delta\eta\equiv\eta_0-\eta_{\rm rec}$. 
We see that the expression for the $\mu$ transfer function $\Delta^\mu_\ell(k)$ contains two terms:
\begin{itemize}
\item First, we have the damping of $\mu$ inhomogeneities from the end of the $\mu$-era to the last-scattering surface, \ie
\begin{equation}
\label{eq:damping_for_forecast}
\mu(\eta_{\rm rec},\vec{k}) = \mu(\eta_f,\vec{k})e^{-k^2\Delta\qD^{-2}}\,\,,
\end{equation}
where $\Delta\qD^{-2}$ is defined in terms of the dissipation scale $\qD$ of $\mu$ inhomogeneities as
\begin{equation}
\Delta\qD^{-2}\equiv\big[\qD^{-2}(z_{\rm rec})-\qD^{-2}(z_{f})\big]\simeq\qD^{-2}(z_{\rm rec})\,\,.
\end{equation}
The expression for $\qD$ as a function of redshift has been derived in Eq.~(4.8) of \cite{Pajer:2013oca}: 
at recombination one has that $\qD^{-2}(z_{\rm rec})\simeq\mpc{0.084}$.
\item Then, there is a spherical Bessel function of projection from the last-scattering surface to the observer at $(\eta_0,\vec{x})$.
\end{itemize}

\begin{figure}[H]
\centering
\includegraphics[width=\columnwidth]{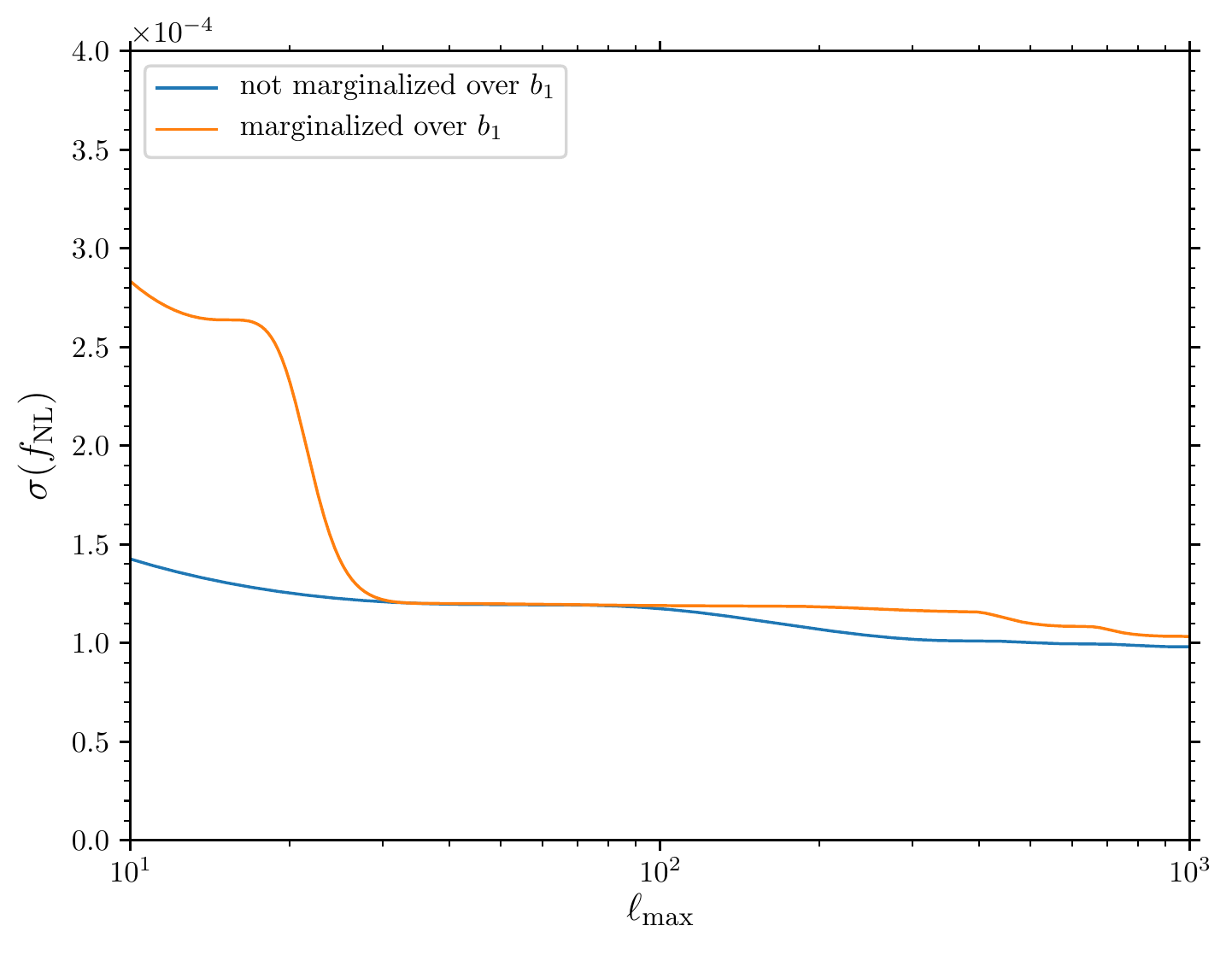} 
\caption{$1\sigma$ detection limits on $\fnl$ for a full-sky cosmic variance-limited experiment as a function of $\ell_{\rm max}$. 
We see that marginalizing over the bias parameter $b_1$ does not significantly affect $\sigma(\fnl)$, 
especially if modes $\ell\gtrsim 20$ are accessible. }
\label{fig:forecast}
\end{figure}

With the expression for the harmonic coefficients of \eq{a_mu_ell_m} at hand, 
we can compute the likelihood in terms of the angular correlators $C^{\mu T}_\ell$, 
$C^{\mu\mu}_\ell$ and $C^{TT}_\ell$. 
The first two are computed in Appendix \ref{app:forecast}, while we use CAMB \cite{Lewis:1999bs} to obtain the latter. 
We start by defining 
\begin{subequations}
\label{eq:like-A}
\begin{align}
&C^{\mu T}_\ell|_{\fnl}\equiv C^{\mu T}_\ell(\fnl=1,b_1=0)\,\,, \label{eq:like-A-1} \\
&C^{\mu T}_\ell|_{b_1}\equiv C^{\mu T}_\ell(\fnl=0,b_1=1)\,\,. \label{eq:like-A-2}
\end{align}
\end{subequations}
Then, if we consider a full-sky cosmic variance-limited experiment, the logarithm of the likelihood reads
\begin{equation}
\label{eq:like-B}
{-2\log\mathcal{L}}=\sum_{\ell=2}^{\ell_{\rm max}} \frac{\big(C^{\mu T}_\ell\big)^2}{\sigma^2_\ell} = 
\sum_{\ell=2}^{\ell_{\rm max}}(2\ell+1)\frac{\big(\fnl C^{\mu T}_\ell|_{\fnl} + b_1 C^{\mu T}_\ell|_{b_1}\big)^2}{C^{\mu\mu}_\ell C^{TT}_\ell}\,\,,
\end{equation}
where we have chosen a zero fiducial value for $\fnl$ and $b_1$ 
(indeed, we are interested in upper limits: moreover $C^{\mu T}_\ell$ is linear in $\fnl$ and $b_1$, 
so taking a non-zero fiducial would not affect the Fisher matrix), 
and we used the fact that, if experimental noise is negligible, the variance of $C^{\mu T}_\ell$ is given by
\begin{equation}
\label{eq:like-C}
\sigma^2_\ell = \big\langle\big(C^{\mu T}_\ell\big)^2\big\rangle - \big\langle C^{\mu T}_\ell\big\rangle^2 = \frac{C^{\mu\mu}_\ell C^{TT}_\ell}{2\ell+1}\,\,.
\end{equation}

It is now straightforward to derive the $1\sigma$ errors on $\fnl$ from \eq{like-B}: the Fisher matrix is defined by 
\begin{equation}
\label{eq:like-D}
F_{ij} ={-\frac{\partial^2\log\mathcal{L}}{\partial p_i\partial p_j}}\qquad\text{($p_1=\fnl$, $p_2=b_1$)}\,\,,
\end{equation}
and the unmarginalized and marginalized $1\sigma$ errors are, respectively, given by
\begin{subequations}
\label{eq:like-E}
\begin{align}
&\sigma(\fnl)|_{\text{unmarg.}}=\frac{1}{\sqrt{F_{11}}}\,\,, \label{eq:like-E-1} \\
&\sigma(\fnl)|_{\text{marg.}}=\sqrt{(F^{-1})_{11}}\,\,. \label{eq:like-E-2}
\end{align}
\end{subequations}
These are shown in Fig.~\ref{fig:forecast}: we see that marginalizing over the bias parameter does 
not degrade the detection limit for $\fnl$, which is of the order of $10^{-4}$ for $\ell_{\rm max}>20$.\footnote{Note that this forecast 
does not include any prior on $b_{1}$ and relies on the shapes only. 
For a realistic prior ($\abs{b_{1}}\lesssim10$) we expect the effect of marginalization to be negligible even at low $\ell$. 
We confirm this in \mbox{Appendix \ref{app:PIXIE}.} } 
This had to be expected, since the effect of spatial curvature on $\mu$ production scales very differently 
in the squeezed limit with respect to local non-Gaussianity. 
Also, we notice that the unmarginalized $1\sigma$ detection limit on $\fnl$ 
is \num{1.2d-4}, which is smaller than the one originally quoted in \cite{Pajer:2012vz}: 
the reason is that there the smearing scale was conservatively taken to be $ k_{\rm D}(z_{f}) $. 
Here, following \cite{Pajer:2013oca}, we instead take it to be $k_{\rm D}(z_{\rm rec})$.

\section{Discussion and conclusions}
\label{sec:conclusions}

\noindent In this work, we studied the prediction for the angular correlator $C^{\mu T}_\ell$ between CMB $\mu$-type spectral distortions and 
temperature anisotropies in single-field, attractor inflation.
We found that the leading term comes from the non-linear effect of a long curvature mode on the \textit{production} of $\mu$ distortions. 
We estimate this effect to be of order $b_1k^2/\mathcal{H}_{f}^2$, where $\mathcal{H}_{f}\simeq\mpc{d-1}$ is the Hubble radius at the end of the $\mu$-era 
and $b_1 = \mathcal{O}(1)$ is a bias parameter. 
Since this contribution shows the scaling with $k_L$ typical of the equilateral shape in the squeezed limit, 
it does not significantly affect searches for local non-Gaussianity as long as observations cover a few tens of $ \ell $. 
To make this more precise, we carry out a forecast for a full-sky cosmic variance-limited experiment. 
We assume that $ b_{1} $ is an unknown parameter which we marginalize over. 
If the experiment has access to modes $\ell\gtrsim 20$, the $1\sigma$ detection limit on $\fnl$ 
remains equal to its unmarginalized value of \num{1.2d-4}. 

We conclude that any constraints on $C^{\mu T}_\ell$ in any foreseeable future 
are indeed direct constraints on the primordial local $f_\text{NL}$: 
any effect coming from late-time evolution has a different $ \ell $ dependence and a negligible amplitude. 
This is in stark contrast with, for example, the CMB squeezed bispectrum, 
where non-primordial effects do lead to an observed local shape even in single-field attractor inflation. 

Note that we have been able to extend our result for the $\mu T$ cross-correlator up to 
very short scales ($\ell\sim{1500}$), because our derivation relies only on the temperature mode being outside the 
Hubble radius at the end of the $\mu$-era, which is a order of magnitude shorter than the Hubble radius at recombination. 
Related to this point, we notice that our work assumed an instantaneous transition from the $\mu$-era to the $y$-era: 
however, we can quickly realize that dropping this assumption would not affect the final results. 
Indeed, assume that $\mu$ production does not stop instantaneously at $z_f$, 
but becomes negligible after some redshift $\tilde{z}_f \lesssim z_f$: 
then, we can just use Weinberg's theorem starting from $\tilde{z}_f$ and proceed in the same way as we discussed in the main text. 
Now the effect of the long mode on $\mu$ production would be of order $k^2/\mathcal{H}^2(\tilde{z}_f) \gtrsim k^2/\mathcal{H}^2(z_f)$. 

We conclude by emphasizing that in this work we have considered only the $\mu T$ angular correlator:
recently, it has been pointed out that the primordial bispectrum can be constrained also by looking at $\mu E$ correlations \cite{Ota:2016mqd}, 
and correlations of temperature and polarization with $y$-type spectral distortions \cite{Emami:2015xqa,Ravenni:2017lgw}. 
More precisely, combining $y$ and $\mu$ distortions offers a powerful way to constrain the running of non-Gaussianity at small scales 
\cite{Emami:2015xqa,Dimastrogiovanni:2016aul,Ravenni:2017lgw}. 
In addition to this, the possibility of using angular three-point functions like $B^{TT\mu}_{\ell_1\ell_2\ell_3}$ has been considered 
as a probe of the primordial trispectrum \cite{Pajer:2012vz,Bartolo:2015fqz,Shiraishi:2016hjd}. 
It would be interesting to extend our results to these observables.

\section*{Acknowledgements}

\noindent It is a pleasure to thank Nicola Bartolo, Marco Celoria, Jens Chluba, Paolo Creminelli, Eiichiro Komatsu, Michele Liguori, 
Mehrdad Mir{\-}ba{\-}bayi, Andrea Ravenni and Fabian Schmidt for useful discussions. 
G. C. acknowledges support from the Starting Grant (ERC-2015-STG 678652) ``GrInflaGal'' of the European Research Council. 
D. vd W. and E. P. are supported by the Delta-ITP consortium, a program of the Netherlands Organization for Scientific
Research (NWO) that is funded by the Dutch Ministry of Education, Culture and Science
(OCW). This work is part of the research programme VIDI with Project No. 680-47-535,
which is (partly) financed by the Netherlands Organisation for Scientific Research (NWO).

\appendix

\section{Window function for \texorpdfstring{$\mu$}{\textbackslash mu} from super-Hubble scales}
\label{app:appendix-1-A}

\noindent In this appendix, we compute the solution to \eq{mu_generation-B} 
for the generation of $\mu$ distortions from damping of acoustic waves 
and discuss what is the leading suppression on large scales (\ie we see how the 
correct general-relativistic calculation leads to the scaling of Eq.~\eqref{eq:window_function-B}). 

In presence of viscous corrections, $\mu$ changes along the fluid lines as
\begin{equation}
\label{eq:mu_generation_reminder}
U^\nu\nabla_\nu\mu = \frac{8t_\gamma}{15(A_n - A_s)}
\bigg({P^\rho_{\hphantom{\rho}(\mu}\nabla_{|\rho|} U_{\nu)}}\nabla^\nu U^\mu 
- \frac{(\nabla_\mu U^\mu)^2}{3}\bigg)\,\,.
\end{equation}
As the right-hand side of the above equation starts at second order in perturbations, 
we assume that also $\mu$ starts at this order. This can be easily seen if we rewrite \eq{mu_generation_reminder} as
\begin{equation}
\label{eq:app-1-A-A}
U^\nu\nabla_\nu\mu = \frac{8t_\gamma}{15(A_n - A_s)}\sigma_{\mu\nu}\sigma^{\mu\nu}\,\,,
\end{equation}
where the anisotropic stress 
$\sigma_{\mu\nu}\equiv P^\rho_{\hphantom{\rho}(\mu}\nabla_{\rho} U_{\nu)} - \frac{\nabla_\rho U^\rho}{3}P_{\mu\nu}$ 
vanishes in a FLRW spacetime. 

Given some foliation $t$ of spacetime, 
with normal $n^\mu$ to the constant-$t$ hypersurfaces, 
one can decompose the fluid velocity $U^\mu$ as (see also \eq{reminder})
\begin{equation}
\label{eq:fluid_velocity_decomposition}
U^\mu = \gamma(n^\mu + v^\mu)\,\,,
\end{equation}
where $v^\mu$ satisfies $h^\mu_{\hphantom{\mu}\nu}v^\nu = v^\mu$ 
($h_{\mu\nu}$ being the projector on constant-$t$ hypersurfaces), 
and the $\gamma$ factor has its usual special relativistic definition. 
It is then straightforward to see that up to first order 
in perturbations the anisotropic stress takes the form 
\begin{equation}
\label{eq:app-1-A-B}
\begin{split}
&\sigma_{\mu\nu} = K_{\mu\nu} + {^{(3)}}\nabla_{(\mu}v_{\nu)} + 2n_{(\mu} K_{\nu)\rho}v^\rho 
- \frac{(K+{^{(3)}}\nabla_\rho v^\rho)h_{\mu\nu}}{3} - \frac{2Kn_{(\mu}v_{\nu)}}{3}\,\,,
\end{split}
\end{equation}
where $K_{\mu\nu} = Hh_{\mu\nu}+\delta K_{\mu\nu}$ and ${^{(3)}}\nabla_\mu$ are, 
respectively, the extrinsic curvature and the covariant derivative on constant-$t$ hypersurfaces. 
Working in Newtonian gauge and dropping vector and tensor modes, \ie
\begin{equation}
\label{eq:newtonian_gauge}
\dif s^2 = -(1+2\Phi)\dif t^2 + a^2(1-2\Psi)\dif\vec{x}^2\,\,,
\end{equation}
simplifies \eq{app-1-A-B} further. 
Indeed, using that $\delta K_{\mu\nu} = -(\dot{\Psi}+H\Phi)h_{\mu\nu}$ at first order in perturbations, 
\ie $K_{\mu\nu} = \frac{K}{3}h_{\mu\nu}$, 
all terms involving the extrinsic curvature cancel. 
Therefore, we remain with 
\begin{equation}
\label{eq:mu_generation_app-B}
\sigma_{\mu\nu} = {^{(3)}}\nabla_{(\mu}v_{\nu)} - \frac{({^{(3)}}\nabla_\rho v^\rho)h_{\mu\nu}}{3}\,\,,
\end{equation}
so that 
\begin{equation}
\label{eq:app-1-A-C}
U^\nu\nabla_\nu\mu = \frac{8t_\gamma}{15(A_n - A_s)}\bigg[{^{(3)}}\nabla_{(\mu}v_{\nu)}{^{(3)}}\nabla^{(\mu}v^{\nu)} 
- \frac{({^{(3)}}\nabla_\mu v^\mu)^2}{3}\bigg]\,\,.
\end{equation}

Integrating \eq{app-1-A-C} in time from $t_i$ to $t_f$ gives the total amount of $\mu$ distortions produced 
by damping of acoustic waves. 
The whole procedure is illustrated in more detail in the Section \ref{ssec:mu_quadratic}: 
here we just want to show that the generation of $\mu$ distortions is suppressed 
(as one would expect) when the wavelength of any of the two modes 
on the right-hand side of \eq{mu_generation_app-B} becomes large. 
For simplicity we consider only the term $({{^{(3)}}\nabla_\mu v^\mu})^2$. 
We can rewrite it in terms of the Newtonian potentials using the shift constraint equation 
(working at linear order in $t_\gamma\times\partial_\mu$, 
we can use the perfect fluid equations). 
Up to second order in perturbations we obtain
\begin{equation}
\label{eq:app-1-A-D}
({{^{(3)}}\nabla_\mu v^\mu})^2 = \frac{4\mpl^4H^2}{a^4(\rho+p)^2}(\partial^2\Phi)^2\,\,,
\end{equation}
so that \eq{app-1-A-C} becomes
\begin{equation}
\label{eq:mu_generation_app-C}
\begin{split}
&\frac{\dot{\mu}}{H}\sim\frac{t_\gamma}{H}\frac{(\partial^2\Phi)^2}{a^2\mathcal{H}^2}\,\,.
\end{split}
\end{equation}
At zeroth order in $t_\gamma$, 
$\Phi$ evolves as $\Phi(\eta,k) = -2\zeta(k)j_1(x)/x$, 
with $x = k\eta/\sqrt{3}$ during radiation dominance. 
Using this solution in \eq{mu_generation_app-C}, we obtain
\begin{equation}
\label{eq:mu_generation_app-D}
\text{$\frac{\dot{\mu}}{H}\sim\frac{1}{\sigma_\text{T}n_ea\cH}\frac{k^2q^2}{\mathcal{H}^2}\zeta(k)\zeta(q)$ for $k,q\to 0\,\,.$}
\end{equation}
We can further simplify this by using the approximate expression 
$k_\text{D}^{-1}\simeq\sqrt{\eta/\sigma_\text{T}n_e a} = \sqrt{1/\sigma_\text{T}n_ea\mathcal{H}}$ 
for the damping scale \cite{Dodelson:2003ft}, 
\ie $k_\text{D}^2\simeq\sigma_\text{T}n_ea\mathcal{H}$. 
Then, we have
\begin{equation}
\label{eq:mu_generation_app-E}
\text{$\frac{\dot{\mu}}{H}\sim\frac{k^2q^2}{\mathcal{H}^2k^2_\text{D}}\zeta(k)\zeta(q)$ for $k,q\to0\,\,.$}
\end{equation}
This makes clear that very long-wavelength modes of $\zeta$ do not dissipate. 
More precisely, the leading contribution to $\dot{\mu}$ will be suppressed: 
taking the short modes to be inside the Hubble radius, 
we see that the $\mu$ production is still suppressed by $k q/k_\text{D}^2$, 
making the generation of $\mu$ distortions from viscosity a process active for $k\gtrsim k_\text{D}$ only. 
This result tells us that, since perturbations longer than the damping scale do not contribute to $\mu$, 
it is possible to treat the effect on $\mu$ production of a long-wavelength 
(that re-enters the Hubble radius at the end of the $\mu$-era) 
$\zeta$ mode as we did in Section \ref{sec:after_mu_era}, 
\ie by an expansion in $k_L/k_S$, with $k_S\sim k_\text{D}(z_f)\simeq\mpc{50}$.

\noindent Writing $v^\mu$ in terms of the velocity potential $v$, 
\ie $v^i = a^{-2}\partial_i v$ in Newtonian gauge \cite{Weinberg:2008zzc}, 
we can rewrite \eq{app-1-A-C} as 
\begin{equation}
\label{eq:app-1-B-A}
\dot{\mu}=\frac{8t_{\gamma}{a^{-4}}}{15(A_{n}-A_{s})}\bigg[
(\partial_{i}\partial_{j}v)^{2}-\frac{(\partial^{2}v)^{2}}{3}\bigg]\,\,.
\end{equation}
We can then insert the linear transfer function for the velocity potential during radiation domination, \ie
\begin{equation}
\label{eq:app-1-B-B}
v(k)=\frac{T_{v}(\eta,k)\zeta(k)}{H}\,\,,
\end{equation}
where we added the Hubble for dimensional consistency. 
This is related to the standard transfer function for the Newtonian potential, 
which is given by
\begin{equation}
\label{eq:app-1-B-C}
\Phi(\eta,k)=-2\zeta(k)\bigg(\frac{\sin{x}-x\cos{x}}{x^{3}}\bigg)\,\,,
\end{equation}
where $x=k\eta/{\sqrt{3}}$. 
Using the $0i$ Einstein equation for scalar perturbations in Newtonian gauge, \ie
\begin{equation}
\label{eq:app-1-B-D}
\frac{(\rho+p)v}{2\mpl^2}=-H\Phi-\dot{\Phi}\,\,,
\end{equation}
we find
\begin{equation}
\label{eq:app-1-B-E}
T_{v}(\eta,k)=\bigg[\bigg(\frac{\sin{x}-x\cos{x}}{x^{3}}\bigg) 
+ \cH^{-1}\partial_{\eta}\bigg(\frac{\sin{x}-x\cos{x}}{x^{3}}\bigg)\bigg]\,\,.
\end{equation}
In momentum space we can write the formula for $\mu$ production as
\begin{equation}
\label{eq:app-1-B-F}
\begin{split}
\mu(\eta_{f},\vec{x}) = \frac{8}{15(A_{s}-A_{n})}&\int_{\vec{k}_1}\int_{\vec{k}_2}\zeta(\vec{k}_{1})\zeta(\vec{k}_{2})\,
e^{i(\vec{k}_{1}+\vec{k}_{2})\cdot\vec{x}} \\
&\times\underbrace{\int^{\infty}_{0}\mathrm{d}z' \left(\frac{t_{\gamma}}{\mathcal{H}}\right)
\bigg(\frac{\left(\vec{k}_{1}\cdot \vec{k}_{2}\right)^{2}-\frac{1}{3}k_{1}^{2}k_{2}^{2}}{\mathcal{H}^{2}}\bigg)\,
T_{v}(z',k_{1})
\,
T_{v}(z',k_{2}) \mathcal{J}_{\mu}(z')}_{\hphantom{W(\vec{k}_1,\vec{k}_2)\,}=\,W(\vec{k}_1,\vec{k}_2)}\,\,.
\end{split}
\end{equation}
We recall that $ \frac{t_{\gamma}}{\cH} \sim \frac{\partial k_\mathrm{D}^{-2}}{\partial z} $, c.f. Eq.~(3) in \cite{Pajer:2012vz}. 
Here we have in mind the exponential suppression of perturbations on scales smaller than the damping scale, 
which restricts the momentum integrals, as in \cite{Pajer:2012vz}.

\section{From the \texorpdfstring{$\mu$}{\textbackslash mu}-era to recombination}
\label{sec:mu_damping}

\noindent In this appendix, we argue that the tight-coupling evolution between the end of the $\mu$-era and recombination \cite{Pajer:2013oca} 
does not lead to a correction to \eq{schematic_correlator}. 
Indeed, viscosity affects only the \textit{inhomogeneities} in the chemical potential, \ie it does not have an effect on a \textit{uniform} $\mu$. 
The presence of a long mode modifies local physics (\textit{e.g.} it changes the local electron density and with it the damping scale), 
but this only leads to couplings of the form $\sim\zeta_L\partial_i\mu_S$: 
we are then in the same situation as in Section \ref{ssec:after_mu_era}, 
where we argued that such terms do not contribute to $\mu T$ cross-correlation. 
In this appendix we put this argument on more formal grounds. 

As long as Compton scattering is efficient, the Boltzmann equation drives the photon distribution towards the equilibrium form. 
That is, the evolution of the chemical potential along the photon geodesics is dictated by
\begin{equation}
\label{eq:tight_coupling_evolution_for_mu}
\frac{1}{E}\frac{\text{D}\mu}{\text{d}\lambda} = \frac{\mu-\mu_0}{t_\gamma}\,\,,
\end{equation}
where $E=-P_\mu U^\mu$ is the photon energy measured by an observer comoving with the fluid, 
and we defined the monopole $\mu_0$ as (we refer to Appendix \ref{app:appendix-3} 
for a more detailed discussion of how integrations over the photon direction can be performed in a general-covariant way) 
\begin{equation}
\label{eq:monopole_definition}
\mu_0(x) = \int\frac{\dif\vers{m}}{4\pi}\mu(x,\vers{m})\,\,,
\end{equation}
with $m^\mu$ being the photon direction as measured by the observer moving along the fluid lines, \ie $P^\mu = E(U^\mu + m^\mu)$. 
We have neglected the quadrupole $\mu_2$ in the collision term: 
indeed, as discussed in \cite{Pajer:2013oca,Ota:2016esq}, it will be suppressed in the tight-coupling regime (together with higher multipoles). 

We then proceed by decomposing $\mu$ as a monopole plus a dipole: 
working non-per{\-}tur{\-}ba{\-}ti{\-}ve{\-}ly in cosmological perturbations, we can write 
\begin{equation}
\label{eq:app_3-B}
\mu = \mu_0 - 3m^\nu P_\nu^{\hphantom{\nu}\rho}\nabla_\rho\mu_1\,\,,
\end{equation}
where $P^\mu_{\hphantom{\mu}\nu}\equiv\delta^\mu_{\hphantom{\mu}\nu} + U^\mu U_\nu$ is the projector on the instantaneous rest frame of the fluid 
and $\mu_0$ and $\mu_1$ are two scalars that we assume to start at second order in the short modes.\footnote{Notice that $\mu_1$ must have 
dimensions of an inverse energy, since $\mu$ is dimensionless: 
one could define $\mu = \mu_0 - 3t_\gamma m^\nu P_\nu^{\hphantom{\nu}\rho}\nabla_\rho\mu_1$ to make $\mu_1$ dimensionless, 
but this is irrelevant for the discussion in this section since we integrate out $\mu_1$.} 

Since $\mu$ does not depend on energy, the left-hand side of \eq{app_3-A} reads as 
\begin{equation}
\label{eq:app_3-C}
\frac{1}{E}\frac{\text{D}\mu}{\text{d}\lambda} = U^\nu\nabla_\nu\mu+m^\nu\nabla_\nu\mu 
+ \frac{P^\rho\nabla_\rho m^\nu}{E} P_{\hphantom{\lambda}\nu}^{\lambda}\frac{\partial\mu}{\partial m^\lambda}\,\,.
\end{equation}
Expanding then the photon geodesic equation, with straightforward manipulations we can see that 
\begin{equation}
\label{eq:app_3-D}
P_{\hphantom{\lambda}\nu}^{\lambda}\frac{P^\rho\nabla_\rho m^\nu}{E} = 
(m^\mu\theta_\mu + m^\mu m^\nu\theta_{\mu\nu})m^\lambda 
- U^\mu\nabla_\mu U^\lambda - m^\mu\theta_\mu^{\hphantom{\mu}\lambda}\,\,,
\end{equation}
where we defined $\theta^\mu\equiv U^\nu\nabla_\nu U^\mu$, $\theta^{\mu\nu}\equiv P^{\mu\rho}\nabla_\rho U^\nu$. 
After plugging \eq{app_3-B} and \eq{app_3-D} in \eq{app_3-C}, 
we can extract two equations for $\mu_0$ and $\mu_1$ by taking moments, 
\ie by integrating the equation in $\int\frac{\dif\vers{m}}{4\pi}(m^\mu m^\nu\dots)$. 
Since we have assumed that $\mu$ is composed by a monopole and a dipole, 
only the first two moments are needed. 
The final result is the system of coupled equations (see Appendix \ref{app:appendix-3})
\begin{subequations}
\label{eq:app_3-F}
\begin{align}
&U^\nu\nabla_\nu\mu_0 + 4\theta^\nu D_\nu\mu_1 - D_\nu D^\nu\mu_1 = 0\,\,, \label{eq:app_3-F-1} \\
&\frac{D^\nu\mu_0}{3} - \frac{4\theta^{(\nu\rho)}D_\rho\mu_1}{5} - \frac{2\theta D^\nu\mu_1}{5} 
+ 3\theta^{\nu\rho}D_\rho\mu_1 - \theta^\nu U^\rho\nabla_\rho\mu_1 
- D^\nu(U^\rho\nabla_\rho\mu_1) = \frac{D^\nu\mu_1}{t_\gamma}\,\,, \label{eq:app_3-F-2}
\end{align}
\end{subequations}
where have denoted the projection of the covariant derivative in the instantaneous rest frame of the fluid by $D_\mu$ 
(\ie $D_\mu\sim P_\mu^{\hphantom{\mu}\nu}\nabla_\nu$). 
Notice that this is different from the covariant derivative on constant-$\eta$ hypersurfaces, $\eta$ being defined by 
\eqsII{adiabatic_mode_in_radiation_domination}{poisson_metric}.


\subsection{Leading order}
\label{ssec:leading_order}

\noindent We start by approaching \eqsI{app_3-F} with the same method that we used to arrive at \eq{mu_observed} 
(the general case will be discussed in Section \ref{ssec:app_subleading_orders}): 
since the two equations are linear in $\mu_0$ and $\mu_1$, 
which we have assumed to start at second order in the short modes, 
we can consider the other tensors like $U^\mu$, $\theta^\mu$, etc. to contain only the long mode $\zeta_L$. 
If we drop all spatial derivatives of the long mode, $\theta^\mu$ vanishes: 
indeed, as we discussed in Section \ref{sec:after_mu_era}, 
the large-scale spatial velocity $v^i_L$ is zero at this order. 
Additionally, $\theta^{\mu\nu}$ is equal to $\theta P^{\mu\nu}/3$ at leading order in $\zeta_L$. 
Then, if we take the three-divergence of \eq{app_3-F-2} 
(again neglecting spatial derivatives of $\zeta_L$), 
we solve it for 
$U^\nu\nabla_\nu(D_\rho D^\rho\mu_1) = D_\rho D^\rho(U^\nu\nabla_\nu\mu_1) + \mathcal{O}(\partial_i\zeta_L)$, 
and plug the result back in the derivative of \eq{app_3-F-1} along $U^\mu$, 
we arrive at
\begin{equation}
\label{eq:app_3-G}
U^\nu\nabla_\nu(U^\rho\nabla_\rho\mu_0) - \frac{D_\nu D^\nu\mu_0}{3} - \bigg(\frac{\theta}{3} -\frac{1}{t_\gamma}\bigg)D_\nu D^\nu\mu_1 = 0\,\,.
\end{equation}
We can then solve algebrically for $D_\nu D^\nu\mu_1$ from \eq{app_3-F-1}, 
plug the result in \eq{app_3-G}, and obtain an equation for the monopole $\mu_0$ alone, \ie
\begin{equation}
\label{eq:app_3-H}
U^\nu\nabla_\nu(U^\rho\nabla_\rho\mu_0) - \frac{D_\nu D^\nu\mu_0}{3} - \bigg(\frac{\theta}{3} - \frac{1}{t_\gamma}\bigg)U^\nu\nabla_\nu\mu_0 = 0\,\,.
\end{equation}
Recalling that at the order we are working at there is no large scale velocity $v^i_L$, 
we can replace $D_\nu D^\nu\mu_0$ by ${^{(3)}\nabla_\nu}{^{(3)}\nabla^\nu}\mu_0$, 
\ie with the covariant spatial derivative on the surfaces of constant time. 
Indeed, ${^{(3)}\nabla_\mu}\sim h_\mu^{\hphantom{\mu}\nu}\nabla_\nu$, 
where $h_{\mu\nu} = g_{\mu\nu} + n_\mu n_\nu$ is the projector on the constant-$\eta$ surfaces. 
Since four-velocity of the fluid can be decomposed as 
\begin{equation}
\label{eq:reminder}
U^\mu = \gamma(n^\mu+v^\mu)\,\,,
\end{equation}
where $\gamma^{-2} = 1-h_{\mu\nu}v^\mu v^\nu$, 
we see that for vanishing $v^\mu$ the two projectors $P_\mu^{\hphantom{\mu}\nu}$ and $h_\mu^{\hphantom{\mu}\nu}$ coincide.
Therefore, we can rewrite \eq{app_3-H} as
\begin{equation}
\label{eq:tight_coupling_evolution_for_mu_monopole}
U^\nu\nabla_\nu(U^\rho\nabla_\rho\mu_0) - \frac{{^{(3)}\nabla_\nu}{^{(3)}\nabla^\nu}\mu_0}{3} 
- \bigg(\frac{\nabla_\mu U^\mu}{3} - \frac{1}{t_\gamma}\bigg)U^\nu\nabla_\nu\mu_0 = 0\,\,.
\end{equation}
The relevant scales in \eq{tight_coupling_evolution_for_mu_monopole} are $H$ 
(from $\nabla_\mu U^\mu/3$), 
$k^2$ (from the three-dimensional Laplacian), 
and $t_\gamma^{-1}\sim k^2_\text{D}/(a\mathcal{H})$ 
(we use the approximate solution $k_\text{D}^2\simeq a\mathcal{H}/t_\gamma$ for the damping scale \cite{Dodelson:2003ft}). 
Up to recombination the comoving damping scale $k_\text{D}^{-1}$ is much shorter than $\mathcal{H}^{-1}$, 
therefore in the bracket multiplying $U^\nu\nabla_\nu\mu_0$ the second term dominates. 
Secondly, ${^{(3)}\nabla_\nu}{^{(3)}\nabla^\nu}\mu_0$ will be subleading with respect to this term if we look at 
correlations with temperature anisotropies that are longer than the damping scale at recombination. 
With these assumptions, we can effectively take the $t_\gamma\to0$ limit of \eq{tight_coupling_evolution_for_mu_monopole}, finding 
\begin{equation}
U^\nu\nabla_\nu\mu_0 = 0\,\,.
\end{equation}
The monopole created after the end of the $\mu$-era is conserved along the fluid lines up to the last-scattering surface. 
This is nothing but \eq{fluid_lines_conservation}, a key assumption in the analysis carried out in Section \ref{sec:after_mu_era}.


\subsection{Subleading orders}
\label{ssec:app_subleading_orders}

\noindent As we are now going to discuss, however, it is also possible to use \eqsI{app_3-F} 
to make a statement more general than just a derivation of \eq{fluid_lines_conservation}. 
Let us follow the results of Section \ref{sec:mu_era} and assume that the initial dipole $\mu_1$ vanishes, 
\ie it is zero at the end of the $\mu$-era.\footnote{The $\mu$-era happens deeply in the tight-coupling regime. 
The assumption of having multipoles beyond the monopole equal to zero during this epoch 
is just the assumption that $\mu$ \textit{production} can be treated within the realm of fluid dynamics, \ie by using \eq{mu_generation-B}. }
Therefore, we see that these \textit{homogeneous} equations have a simple solution if the initial monopole is uniform. 
Indeed, for these initial conditions, the solution is 
\begin{equation}
\mu_1\equiv 0 \qquad \text{and} \qquad \mu_0 = \text{const.}\,\,.
\end{equation} 
In words, \textit{only the inhomogeneities of the monopole evolve after the end of the $\mu$-era}. 
We regard this as a generalization of the result of \cite{Pajer:2013oca}, 
that showed how the linear evolution up to recombination is a damping of the spatial fluctuations in the monopole. 

One might wonder what happens if we relax the assumption of having the chemical potential \textit{at recombination} consist of only a monopole. 
We can see that our conclusions will not be changed in the following way: 
the free-streaming solution will still be that of \eq{solution_to_collisionless_boltzmann}, 
since \eq{collisionless_boltzmann} must be satisfied for every value of the photon energy and both $\mu_0$ and $\mu_1$ in \eq{app_3-B} do not depend on $E$. 
Then, let us consider the solution of \eq{solution_to_collisionless_boltzmann} at zeroth order in the long modes 
(this is enough, since any ``projection effect'' interaction will be built from powers of $\zeta_L$ multiplying the zeroth-order solution, 
as we discussed in Section \ref{sec:after_mu_era}).
We see that, using the fact that at this order $m^\mu$ is conserved along the photon geodesics 
(lensing begins at first order in perturbations), the contribution of the dipole to the full chemical potential at the observer's point is of the form 
$\mu\sim\vers{n}\cdot\vec{\nabla}\!_{\vers{n}}\mu_1$. This contribution vanishes once we average over the short modes, 
since $\braket{\mu_1}$ does not depend on position (the proof is the same as the one in Appendix \ref{app:appendix-2}).\footnote{Equivalently, 
we cannot have a inhomogeneous average dipole $\braket{\mu_1}$ if statistical isotropy is satisfied. 
This would amount to have a preferred vector $P_\nu^{\hphantom{\nu}\rho}\nabla_\rho\braket{\mu_1}$ (see \eq{app_3-B}).}

\section{Cancellation of \texorpdfstring{$f_\mathrm{NL} = 1-n_\mathrm{s}$}{f\_\{NL\} = 1-n\_s} in global coordinates}
\label{app:nsmin1}

\noindent In this appendix we comment on the relation of our result to Maldacena's consistency condition for the squeezed bispectrum \cite{Maldacena:2002vr}, 
which suggests a local $\fnl$ of $1-n_{\rm s}$. In our result no such term appears. 
The reason is related to the discussion in the conclusions of \cite{Pajer:2012vz}: 
Maldacena's result is obtained in particular coordinates, which are not necessarily well-suited to compute observables. 
We stress that our derivation in the main text is more complete, 
coordinate-independent and exact up to the corrections discussed in Section \ref{sec:leading_effect}.

\subsection*{\texorpdfstring{$\braket{\mu T}$}{<\textbackslash muT>} in global coordinates}

\noindent Using global coordinates in $\zeta$ gauge, the three-point function in single-field cosmology is given by \cite{Maldacena:2002vr}
\begin{equation}
\braket{\zeta(\vec{q}\to 0)\zeta(\vec{k})\zeta(-\vec{k})}' = {-\frac{\dif\log k^{3}P_{\zeta}(k)}{\dif\log k}}P_{\zeta}(q)P_{\zeta}(k)\,\,. 
\end{equation}
If we combine this with the sub-Hubble expression for $\mu$ production, 
one naively concludes that there is a nonzero contribution to $\braket{\mu T}$ given by
\begin{equation}
\begin{split}
\braket{\mu (\vec{q}) T(-\vec{q})}'&\sim -(n_\mathrm{s}-1)\int_{k_{\mathrm{D}}(z_f)}^{k_{\mathrm{D}}(z_i)}\dif\log k\,\Delta_\zeta^{2}(k)P_{\zeta}(q) \\
&= \bigg\langle{\int_{k_{\mathrm{D}}(z_f)}^{k_{\mathrm{D}}(z_i)}\dif\log k\,\Delta^{2}_{\zeta,{\rm G}}\big(k(1-\zeta_{L})\big)\zeta(q)}\bigg\rangle\,\,,
\end{split}
\end{equation}
where the subscript ``$\mathrm{G}$'' denotes the fact that this quantity is uncorrelated with the long mode. 
This last equality is the crucial observation in this context: 
Maldacena's result can be interpreted as a local shift of coordinates in the presence of a long mode, 
see \cite{Weinberg:2008zzc,Mirbabayi:2014hda}. 
Since $\mu$ is being created during this era, it is sensitive to a change of coordinates $\mu(t,\vec{x})$. 
This naive calculation is wrong, because it does not consistently use the same coordinates to describe every scale present in the computation. 
Namely, a local change of coordinates (simply a spatial dilation in $\zeta$ gauge \cite{Maldacena:2002vr,Creminelli:2004yq}) 
affects all local physics in the same way. 
In particular, the damping scale at the beginning and end of the $\mu$-era is also slightly modified. 
Thus, the proper expression for the creation of $\mu$ in global coordinates is
\begin{equation}
\big\langle\langle\mu\rangle_{\zeta_{L}} (\vec{q}) T(-\vec{q})\big\rangle^{\prime}\propto\bigg\langle
\int_{k_{\mathrm{D}}(z_f)\times(1-\zeta_{L})}^{k_{\mathrm{D}}(z_i)\times(1-\zeta_{L})}
\dif\log{\big(k(1-\zeta_{L})\big)}\,\Delta^{2}_{\zeta,{\rm G}}\big(k(1-\zeta_{L})\big)\zeta(q)\bigg\rangle=0\,\,,
\end{equation}
where $\langle\mu\rangle_{\zeta_{L}} (\vec{q})$ denotes the expectation value of $\mu$ in the presence of a constant long mode. 
Substituting variables, we can simply remove any dependence of the short modes on the long modes, such that, after the $\mu$-era, 
the correlation is zero indeed. The intuition is that regardless of which coordinates one uses, the total physical duration of the $\mu$-era is always the same.

\section{Average \texorpdfstring{$\mu$}{\textbackslash mu} distortions in the sky}
\label{app:appendix-2}

\noindent In this appendix, we provide a more detailed proof of \eq{schematic_correlator}. 
As we have seen in Section \ref{sec:after_mu_era}, we can have two effects that involve the long-wavelength mode. 
First of all, its presence affects the relation between the direction of observation and the physical position at recombination. 
Then, it affects the observed $\mu$ anisotropies through the second term of \eq{mu_observed}, 
\ie the effect of the long mode on the short modes at the end of the $\mu$-era as obtained from Weinberg's theorem, \eq{mu_transformation}. 
Both effects involve derivatives of $\mu_S(\eta_f,\vers{n}(\eta_0 - \eta_\text{rec}))$ with respect to $\vers{n}$ 
(as can be checked by expanding $\mu_S(\eta_f,\vec{x}_\text{rec})$ at leading order in $\zeta_L$): 
therefore, if the ensemble average of $\mu_S(\eta_f,\vers{n}(\eta_0 - \eta_\text{rec}))$ is independent on the direction of observation, 
the final $C^{\mu T}_\ell$ will vanish. 
This is straightforward to see: 
indeed, from Section \ref{sec:mu_era} we know that $\mu(\eta_f,\vec{x})$ can be written as 
(we will drop the subscript $S$ for simplicity) 
\begin{equation}
\label{eq:app_2}
\mu(\eta_f,\vec{x}) = \int_{\vec{k}_1}\int_{\vec{k}_2}\zeta(\vec{k}_1)\zeta(\vec{k}_2)
W(\vec{k}_1,\vec{k}_2)e^{i(\vec{k}_1+\vec{k}_2)\cdot\vec{x}}\,\,.
\end{equation}
The free-streaming solution at zeroth order in $\zeta_L$ just corresponds to replacing $\vec{x}$ with $\vec{x}_0+\vers{n}(\eta_0 - \eta_\text{rec})$, 
as seen in \eq{x_rec_vs_x_obs}: 
therefore, when taking the ensemble average, the two momenta $\vec{k}_1$ and $\vec{k}_2$ 
are forced to be equal and opposite and any dependence on $\vers{n}$ drops out.

\section{Moments of the Boltzmann equation}
\label{app:appendix-3}

\noindent In this appendix, we collect some details of the calculations that lead to \eqsI{app_3-F}. 
We recall that our starting equation for the evolution of the chemical potential is \eq{tight_coupling_evolution_for_mu}, \ie
\begin{equation}
\label{eq:app_3-A}
\frac{1}{E}\frac{\text{D}\mu}{\text{d}\lambda} = \frac{\mu-\mu_0}{t_\gamma}\,\,.
\end{equation}
In the above equation, $E$ is the photon energy measured by the observer $U^\mu$ comoving with the fluid, 
and we assume that $\mu$ is only a function of the photon direction $m^\mu$, defined by $P^\mu = E(U^\mu+m^\mu)$. 

After expanding the derivative $\frac{\text{D}}{\text{d}\lambda}$ along the photon geodesics, 
as detailed in Section \ref{sec:mu_damping}, we can take moments of \eq{app_3-A}. 
As shown in \cite{Misner:1974qy} (see for example its Chapter 22) 
and Appendix A of \cite{Senatore:2008vi}, given a function $F=F(P^\mu,U^\nu,g_{\rho\sigma})$ 
we can integrate it over $P^\mu$ using the Lorentz-invariant measure $\frac{\dif^3P}{E(\vec{P})}$, 
where a local Lorentz frame in which $U^\mu = \delta^\mu_0$ is used to define the components of $\vec{P}$, 
and the positive-energy solution of $P_\mu P^\mu = -m^2$ is selected in the relation $E = E(\vec{P})$. 
For photons, this amounts to writing
\begin{subequations}
\label{eq:measure}
\begin{align}
&m^\mu = (0,\sin\theta\cos\phi,\sin\theta\sin\phi,\cos\theta)\,\,, \label{eq:measure-1} \\
&E=\abs{\vec{P}}\equiv P\,\,, \label{eq:measure-2} \\
&\frac{\dif^3P}{E(\vec{P})} = P\dif P\,\sin\theta\dif\theta\,\dif\phi\,\,. \label{eq:measure-3}
\end{align}
\end{subequations}
In the case that the function $F$ does not depend explicitly on $E$, as it is the case for both the left-hand and right-hand sides of \eq{app_3-A}, 
we can forget about the ``radial'' integration and write $\int\dif\phi\,\dif\theta\sin\theta\equiv\int\dif\vers{m}$.\footnote{Equivalently, 
integrating over $P$ will give the same overall factor on both sides of \eq{app_3-A}.}
With this result at hand, we also see that the definition of the monopole $\mu_0$ as in \eq{monopole_definition} is now made rigorous. 

On the left-hand side of \eq{app_3-A}, coming from the expansion of $\frac{\text{D}}{\text{d}\lambda}$, 
there will be many terms involving tensors orthogonal to the fluid velocity contracted with different powers of $m^\mu$: 
here we collect some useful results that are needed to get the zeroth and first moments. 
Given a vector $V^\mu$ and a tensor $M^{\mu\nu}$, both orthogonal to $U^\mu$ (so that in the local Lorentz frame they will have only spatial components), 
we have that
\begin{subequations}
\label{eq:app_3-E}
\begin{align}
&\int\frac{\dif\vers{m}}{4\pi}\,m^\mu V_\mu = 0\,\,, \label{eq:app_3-E-1} \\
&\int\frac{\dif\vers{m}}{4\pi}\,m^\mu m^\nu M_{\mu\nu} = \frac{M_\mu^{\hphantom{\mu}\mu}}{3}\,\,, \label{eq:app_3-E-2} \\
&\int\frac{\dif\vers{m}}{4\pi}\,m^\mu m^\nu M_{\mu\nu}\,m^\rho V_\rho = 0\,\,, \label{eq:app_3-E-3} \\
&\int\frac{\dif\vers{m}}{4\pi}\,m^\mu\,m^\nu V_\nu = \frac{V^\mu}{3}\,\,, \label{eq:app_3-E-4} \\
&\int\frac{\dif\vers{m}}{4\pi}\,m^\mu\,m^\nu m^\rho M_{\nu\rho} = 0\,\,, \label{eq:app_3-E-5} \\
&\int\frac{\dif\vers{m}}{4\pi}\,m^\mu\,m^\nu m^\rho M_{\nu\rho}\,m^\lambda V_\lambda = \frac{2M^{(\mu\nu)}V_\nu 
+ M_\nu^{\hphantom{\nu}\nu}V^\mu}{15}\,\,. \label{eq:app_3-E-6} 
\end{align}
\end{subequations}

\section{Details of the Fisher forecast}
\label{app:forecast}

\noindent In this appendix, we collect some useful results that are needed to carry out the Fisher forecast of Section \ref{sec:forecast}. 
As discussed in the main text, the expression for the decomposition of $\mu$ on the sky in spherical harmonics reads
\begin{equation}
\label{eq:forecast_app-A}
a^\mu_{\ell m}(\eta_0,\vec{x}) = 4\pi\, i^{-\ell}\int_{\vec{k}}e^{i\vec{k}\cdot\vec{x}}\mu(\eta_f,\vec{k})\Delta^\mu_\ell(k)Y^\ast_{\ell m}(\vers{k})\,\,,
\end{equation}
where 
\begin{equation}
\label{eq:forecast_app-B}
\Delta^\mu_\ell(k)=e^{-\frac{k^2}{\qD^2(z_{\rm rec})}}j_\ell(k\Delta\eta)\,\,,
\end{equation}
with $\Delta\eta\equiv\eta_0-\eta_{\rm rec}$ and $\qD^2(z_{\rm rec})\simeq\mpc{0.084}$. 

With this expression we can then readily compute the $\mu T$ and the $\mu\mu$ angular correlators. 
We start by computing the two correlators $C^{\mu T}_\ell|_{\fnl}$ and $C^{\mu T}_\ell|_{b_1}$, 
which we have defined in \eq{like-A}. 
The general expression for the angular correlator is given by 
\begin{equation}
\label{eq:forecast_app-C}
\braket{a^\mu_{\ell m}(a^T_{\ell'm'})^\ast} = (4\pi)^2i^{-\ell+\ell'}\int_{\vec{a}}\int_{\vec{b}}e^{i(\vec{a}-\vec{b})\cdot\vec{x}}
\braket{\mu(\eta_f,\vec{a})\zeta(-\vec{b})}\Delta^\mu_\ell(a)\Delta^T_{\ell'}(b)Y^\ast_{\ell m}(\vers{a})Y_{\ell'm'}(\vers{b})\,\,.
\end{equation}
In the squeezed limit we can write the ensemble average $\braket{\mu(\eta_f,\vec{a})\zeta(-\vec{b})}$ as 
\begin{equation}
\label{eq:forecast_app-D}
\braket{\mu(\eta_f,\vec{a})\zeta(-\vec{b})}' = \braket{\mu(\eta_f,\vec{x})}P_\zeta(b)\bigg[{\frac{12\fnl}{5}} + \frac{b_1 b^2}{\cH_f^2}\bigg]\,\,,
\end{equation}
where $\braket{\mu(\eta_f,\vec{x})}$ is obtained by taking the ensemble average of, \textit{e.g.}, \eq{window_function-A}. 
\eq{forecast_app-D}, then, leads to
\begin{subequations}
\label{eq:forecast_app-E}
\begin{align}
&C^{\mu T}_\ell|_{\fnl} = \frac{24\braket{\mu}}{5\pi}\int_0^{+\infty}\dif b\,b^2P_\zeta(b)\Delta^\mu_\ell(b)\Delta^T_\ell(b)\,\,, \label{eq:forecast_app-E-1} \\
&C^{\mu T}_\ell|_{b_1} = \frac{2\braket{\mu}}{\pi\mathcal{H}^2_f}\int_0^{+\infty}\dif b\,b^4P_\zeta(b)\Delta^\mu_\ell(b)\Delta^T_\ell(b)\,\,, \label{eq:forecast_app-E-2}
\end{align}
\end{subequations}
where we called $\braket{\mu}\equiv\braket{\mu(\eta_f,\vec{x})}$ for simplicity of notation. 

Then, we move to the computation of the $\mu\mu$ angular power spectrum. 
The steps of the calculation are similar to those above, 
\textit{i.e.} we write 
\begin{equation}
\label{eq:forecast_app-E-bis}
\braket{a^\mu_{\ell m}(a^\mu_{\ell'm'})^\ast} = (4\pi)^2i^{-\ell+\ell'}\int_{\vec{a}}\int_{\vec{b}}e^{i(\vec{a}-\vec{b})\cdot\vec{x}}
\braket{\mu(\eta_f,\vec{a})\mu(\eta_f,-\vec{b})}\Delta^\mu_\ell(a)\Delta^\mu_{\ell'}(b)Y^\ast_{\ell m}(\vers{a})Y_{\ell'm'}(\vers{b})\,\,.
\end{equation}
Now, however, we need to compute the Gaussian contribution to the ensemble average $\braket{\mu(\eta_f,\vec{a})\mu(\eta_f,-\vec{b})}$. 
Using Wick's theorem and working in the squeezed limit, it is straightforward to see that it is made up of a 
``connected'' and a ``disconnected'' contribution: 
\begin{equation}
\label{eq:forecast_app-F}
\braket{\mu(\eta_f,\vec{a})\mu(\eta_f,-\vec{b})} = (2\pi)^6\delta^{(3)}(\vec{a})\delta^{(3)}(-\vec{b})\braket{\mu}^2 + 2F(2\pi)^3\delta^{(3)}(\vec{a}-\vec{b})\,\,,
\end{equation}
where $F$ is given by
\begin{equation}
\label{eq:forecast_app-G}
F=\frac{1}{2\pi^2}\int_0^{+\infty}\dif k\,k^2P^2_\zeta(k)W^2(k,k)\,\,.
\end{equation}
The disconnected one does not contribute to the angular correlator,\footnote{Indeed, 
it forces both transfer functions in \eq{forecast_app-E-bis} 
to be evaluated at zero momentum, where they vanish for $\ell>0$ (since $j_\ell(x)\sim x^\ell$ for $x\to 0$). } 
while the connected one gives
\begin{equation}
\label{eq:forecast_app-H}
C^{\mu\mu}_\ell = \frac{4F}{\pi}\int_0^{+\infty}\dif b\,b^2\big[\Delta^\mu_\ell(b)\big]^2\,\,.
\end{equation}

\section{Fisher forecast for a PIXIE-like experiment}
\label{app:PIXIE}

\noindent In this appendix we carry out a forecast for a PIXIE-like experiment in a similar way to that of Section \ref{sec:forecast}. 
We assume isotropic white noise, a $1\sigma$ uncertainty on the $\mu$ monopole of $10^{-8}$ and a Gaussian beam with a
full-width-at-half-maximum $\theta_{\rm b}=\ang{1.6}$ \cite{Pajer:2012vz,Ganc:2012ae}. Correspondingly, we have that
\begin{equation}
\label{eq:noise_PS_PIXIE}
C^{\mu\mu,{\rm N}}_\ell = 4\pi\times10^{-16}\,e^{\frac{\ell^2\theta_{\rm b}^2}{8\log 2}} = 4\pi\times10^{-16}\,e^{\frac{\ell^2}{84^2}}\,\,,
\end{equation}
which is much larger than $C^{\mu\mu}_\ell$ \cite{Pajer:2012vz}. Therefore, \eq{like-B} becomes
\begin{equation}
\label{eq:like_PIXIE}
{-2\log\mathcal{L}}=\sum_{\ell=2}^{\ell_{\rm max}}(2\ell+1)\frac{\big(\fnl C^{\mu T}_\ell|_{\fnl} + b_1 C^{\mu T}_\ell|_{b_1}\big)^2}{C^{\mu\mu,{\rm N}}_\ell C^{TT}_\ell}\,\,.
\end{equation}
We can also studying what happens if we add a physically-motivated prior on $b_1$. 
We consider a (very conservative) Gaussian prior centered around $b_1=0$ and with $\sigma_{b_1}=10$: 
the log-likelihood of \eq{like_PIXIE}, then, becomes
\begin{equation}
\label{eq:like_PIXIE_prior}
{-2\log\mathcal{L}}=\sum_{\ell=2}^{\ell_{\rm max}}(2\ell+1)\frac{\big(\fnl C^{\mu T}_\ell|_{\fnl} + b_1 C^{\mu T}_\ell|_{b_1}\big)^2}{C^{\mu\mu,{\rm N}}_\ell C^{TT}_\ell} 
+ \frac{b_1^2}{\sigma^2_{b_1}}\,\,.
\end{equation}

\begin{figure}[H]
\centering
\begin{tabular}{c}
\includegraphics[width=0.75\columnwidth]{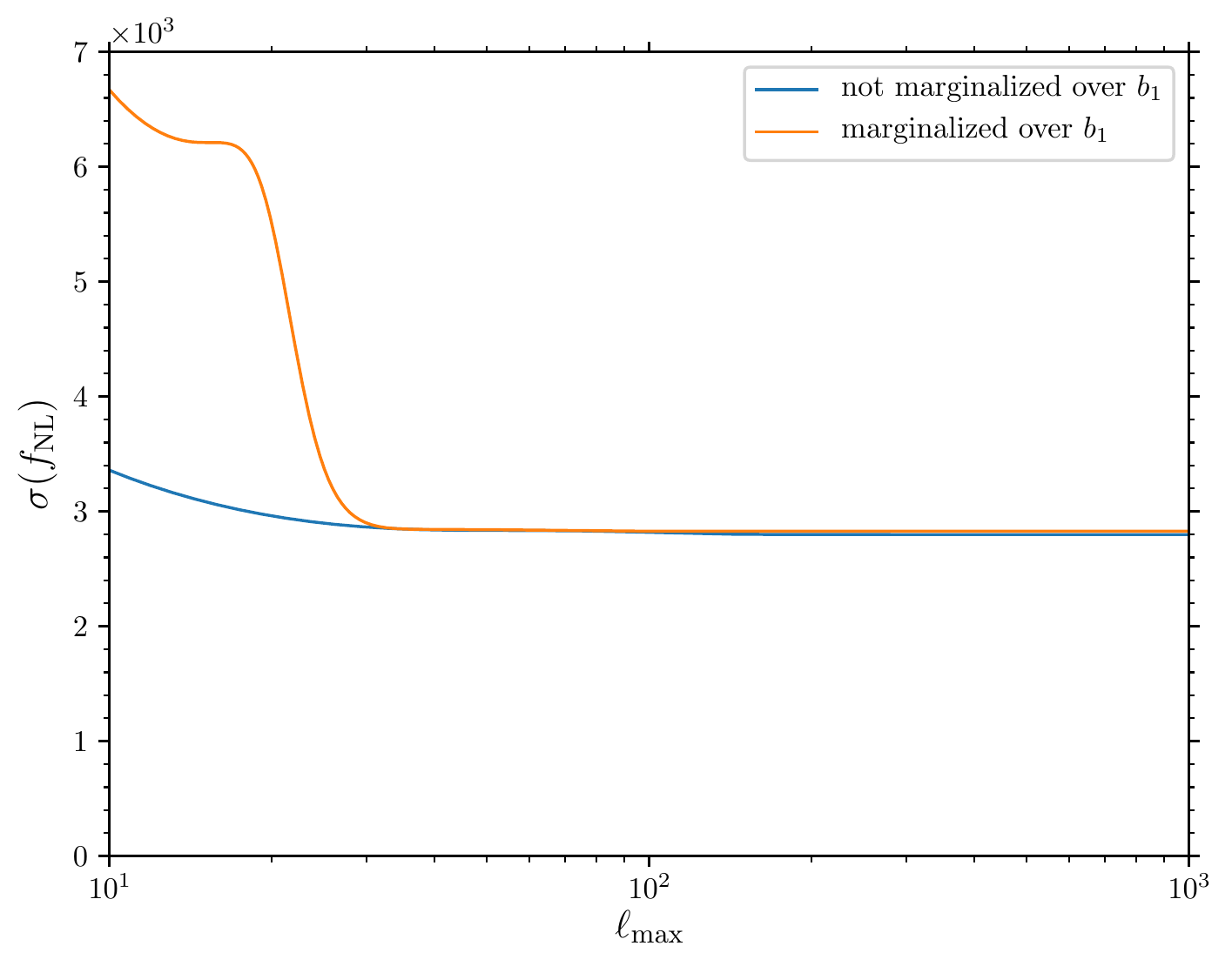} \\
\includegraphics[width=0.75\columnwidth]{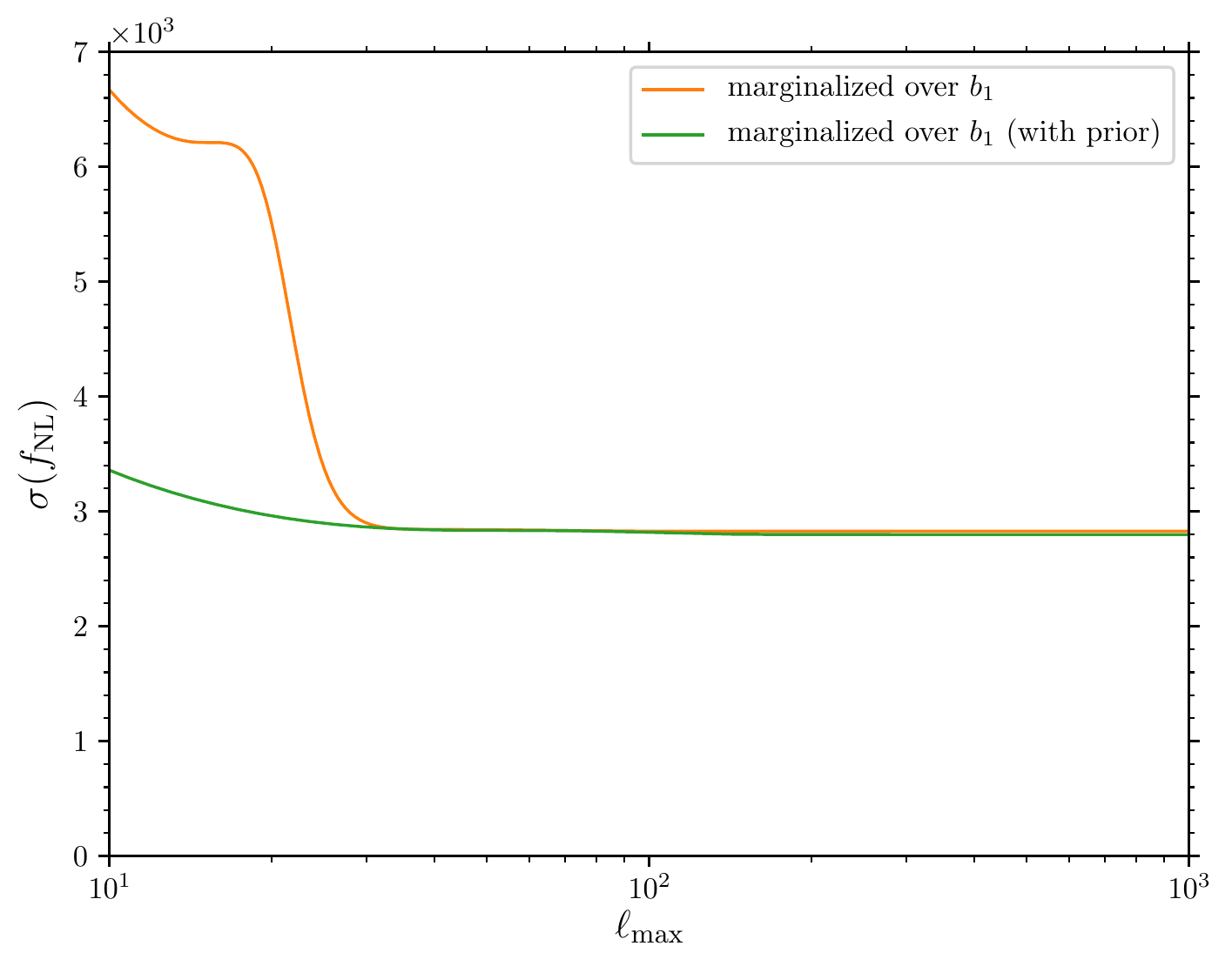}
\end{tabular}
\caption{$1\sigma$ detection limits on $\fnl$ for a PIXIE-like experiment as a function of $\ell_{\rm max}$. 
Top panel: same as Fig.~\ref{fig:forecast}. Bottom panel: added a Gaussian prior on $b_1$. }
\label{fig:PIXIE_forecast}
\end{figure}

The results of the forecast are shown in Fig.~\ref{fig:PIXIE_forecast}. From the top panel we 
see that marginalizing over the bias parameter $b_1$ without any prior 
does not affect $\sigma(\fnl)$ since PIXIE would be able to access modes with $\ell\gtrsim 20$. 
We also see that there is no improvement if we go to $\ell_{\rm max}$ larger than $\mathcal{O}(100)$ since 
PIXIE does not have access to those scales, as shown in \eq{noise_PS_PIXIE}. 
In the bottom panel we show what happens if we include a prior on $b_1$: 
we see that, in this case, even if we stop at scales $\ell_{\rm max} \lesssim 20$ 
the different scale dependence of the two signals can be resolved. Consequently, 
marginalizing over $b_1$ does not affect $\sigma(\fnl)$. 

This confirms that the non-primordial effects are completely orthogonal to local non-Gaussianity and will 
not bias future constraints on $\fnl$ in any way.

\clearpage


\bibliographystyle{utphys}
\bibliography{refs}

\end{document}